\documentclass[12pt,english]{article}

\usepackage[T1]{fontenc}
\usepackage[utf8]{inputenc}   
\usepackage{babel}
\usepackage{microtype}

\usepackage{amsmath}
\usepackage{amssymb}
\usepackage{amsthm}
\usepackage{dsfont}

\usepackage{textcomp}

\usepackage{newtxtext}
\usepackage[cmintegrals,varg]{newtxmath} 

\usepackage{graphicx}
\usepackage{tikz}
\usetikzlibrary{decorations.markings}
\usetikzlibrary{arrows.meta} 

\theoremstyle{plain} 
\newtheorem{theorem}{Theorem}

\newtheorem{lemma}{Lemma}
\newtheorem{corollary}{Corollary}

\theoremstyle{definition} 

\newtheorem{example}{Example} 

\theoremstyle{remark} 
\newtheorem{remark}{Remark}
\newtheorem*{remark*}{Remark} 

\usepackage{makecell}
\usepackage{arydshln}
\usepackage{tabularx}
\usepackage{booktabs}
\newcolumntype{Y}{>{\centering\arraybackslash}X}

\usepackage{siunitx}
\usepackage{comment}
\usepackage{listings}
\usepackage{xcolor}

\definecolor{verylightgray}{gray}{0.95}
\definecolor{CommentGreen}{rgb}{0.2,0.54,0.1}

\usepackage[authoryear]{natbib}

\usepackage{geometry}
\usepackage{setspace}
\usepackage[colorlinks,
            linkcolor=blue,
            anchorcolor=blue,
            citecolor=blue]{hyperref}

\makeatother

\title{Moments by Integrating the Moment-Generating Function\thanks{Corresponding author: Chen Tong, Email: tongchen@xmu.edu.cn. We are
grateful to Christian Berg, Markus Bibinger, Simon Broda, Raymond
Kan, Joseph Romano, and Michael Wolf for helpful and valuable comments.
Chen Tong acknowledges financial support from the National Natural Science Foundation of China (72301227) and the Fujian Provincial Natural Science Foundation of China (2025J08008).}}

\author{Peter Reinhard Hansen$^{\mathsection}$ \ \  and \ \ Chen Tong$^{\ddagger}$
\\[0.1cm] \small $^{\mathsection}$Department of Economics, University of North Carolina at Chapel Hill
\\[-2mm] \small $^{\ddagger}$School of Economics, Xiamen University}

\date{\small \today\vspace{-10mm}}

\begin{document}

\maketitle

\begin{abstract}
We introduce a general integral framework for computing fractional, complex, absolute, and logarithmic moments from the moment-generating function (MGF) under explicit regularity conditions. By evaluating a complex extension of the MGF along a vertical contour, we obtain exact integral expressions that bypass the need for explicit probability densities and high-order derivatives. We establish conditions for negative fractional moments using the symmetric Cauchy principal value, including the requirement that the distribution have no point mass at the centering point. We demonstrate the theoretical scope and computational practicality of the framework through applications to the normal-inverse Gaussian distribution and a semicontinuous compound Poisson-Gamma distribution. In the latter case, the framework handles point masses at the boundary by evaluating conditional fractional moments.
\end{abstract}

{\small\textit{{\noindent}Keywords:}}{\small{} Moments, Fractional
Moments, Moment-Generating Function.}{\small\par}

\newpage

\section{Introduction}

The computation of fractional, absolute, and complex moments is a persistent analytical challenge in applied probability and mathematical statistics, particularly for distributions that lack tractable probability densities. In this paper, we introduce a contour integral framework for obtaining a wide variety of moments for random variables admitting the relevant one-sided or two-sided exponential moments, with additional zero-mass and integrability conditions for negative and logarithmic moments. The proposed method provides exact integral representations that avoid explicit probability densities and high-order derivatives, and it is computationally efficient in the examples considered below.

It is a fundamental result that the $k$-th integer moment of a random variable $X$, with MGF $M_{X}(s)=\mathbb{E}[e^{sX}]$, is given by $\mathbb{E}[X^{k}]=M_{X}^{(k)}(0)$. However, non-integer moments and fractional absolute moments, $\mathbb{E}|X|^{r}$ for $r\notin\mathbb{N}_{0}$, do not admit such straightforward evaluations. Existing methods typically rely on complex integrals involving successive derivatives of the characteristic function (CF) or fractional calculus.\footnote{\citet{Kawata1972} provides an expression for $\mathbb{E}|X|^{r}$ that involves several derivatives of the CF, and \citet{Laue1980} provides an expression based on fractional derivatives; see also \citet{Wolfe:1975}, \citet{SamkoKilbasMarichev:1993}, \citet{MatsuiPawlas2016}, and \citet{Tomovski2022}. For fractional moments of non-negative variables, similar expressions were derived in \citet{CressieDavisFolksPolicello:1981}, \citet{CressieBorkent1986}, \citet{Jones1987}, \citet{SchurgerK:2002}, and \citet{Meng:2005}.} 
In this paper, we derive new integral expressions for computing fractional, absolute, central, and logarithmic moments that entirely bypass the need for probability densities or MGF derivatives.

Our approach is grounded in the Fourier-Laplace transform. By analytically continuing the MGF into the complex plane, $M_{X}(s+it)$, we evaluate the transform along vertical contours. This complex moment-generating function (CMGF) framework is flexible enough to handle complex powers, $r\in\mathbb{C}$, which frequently arise in analytic number theory, statistical mechanics, and extreme value theory. Crucially, we formalize the topological conditions required to extend these representations to negative fractional moments, $-1<\operatorname{Re}(r)\leq0$, utilizing the symmetric Cauchy principal value and establishing necessary zero-mass constraints.

The advantages of the framework become evident when applied to analytically intractable distributions. For instance, existing expressions for fractional moments of the normal-inverse Gaussian (NIG) distribution involve infinite series of modified Bessel functions of the second kind, making the resulting formulas algebraically cumbersome and less convenient for numerical evaluation. The CMGF framework replaces these infinite series with a single contour integral involving the closed-form MGF. Furthermore, the method applies to distributions lacking closed-form densities, such as the semi-continuous compound Poisson-Gamma (Tweedie) distribution, while addressing the topological complication introduced by a point mass at zero.

We are not the first to leverage the MGF in integral expressions. \citet{Meng:2005} derived integral expressions for $\mathbb{E}[X^{a}/Y^{b}]$ involving the joint MGF $M_{X,Y}(s_{1},s_{2})$ and its derivatives, noting the potential of combining characteristic and moment-generating functions in his epilogue. For positive-part moments, our expressions build upon and generalize the integral identities established by \citet{Pinelis:2011}, expanding the domain to include negative and complex moments. Furthermore, we connect this probability framework to recent advances in special functions. 
Existing integral representations for fractional moments are closely related but typically more specialized. For variables on the real line, \citet{Kawata1972} and \citet{Laue1980} express fractional absolute moments through characteristic functions and their derivatives, with the derivative order depending on the moment order. For non-negative random variables, \citet{SchurgerK:2002} and \citet{CressieBorkent1986} provide Laplace-transform or MGF-based formulas, again involving derivatives or one-sided support restrictions; \citet{SchurgerK:2002} also gives a particularly simple Laplace-transform representation for negative moments of strictly positive variables. Our contribution is complementary but distinct: by evaluating the MGF on a vertical contour and using the vanishing identity, we obtain derivative-free formulas that apply to absolute moments, positive-part moments, signed integer moments, logarithmic moments, and complex powers in a single framework. Section \ref{sec:theory} develops these identities, while the Supplementary Material summarizes the closest existing formulas for comparison.

While many existing moment identities based on Fourier-Laplace transforms are derived via differentiation of the MGF/characteristic function or are stated in their simplest form for one-sided laws, our approach starts from a unified absolute-moment identity and combines it with a vanishing-identity step that cancels the additional terms arising in the Fourier-Laplace representation. This yields moment formulas that extend to negative and complex powers under symmetric Cauchy principal value conditions. From a computational standpoint, the resulting expressions are derivative-free and numerically stable, and they accommodate semicontinuous laws in a natural way by working with conditional moments on the positive component.

The remainder of this paper is organized as follows. Section \ref{sec:theory} develops the general theory of complex and fractional moments. 
Theorem \ref{thm:factional-abs-moments} establishes the fundamental representation for absolute moments. Corollary \ref{cor:tilted_moments} extends this to exponentially tilted fractional moments, and Corollary \ref{cor:log_moments} extracts logarithmic moments by differentiating with respect to the moment order. Theorem \ref{thm:pos-part-moments} presents positive-part moments, and Theorem \ref{thm:integer-moments} recovers integer moments without differentiating the MGF. Finally, Theorem \ref{thm:parabolic-positive} uses the reciprocal-Gamma identity from the Parabolic Mellin Transform to obtain a parabolic-contour representation for strictly positive random variables with Gaussian decay.
Section 3 applies this theoretical framework to resolve the analytical bottlenecks of the normal-inverse Gaussian and compound Poisson-Gamma distributions. Section 4 concludes.

\section{Integral Representations via the Complex Moment-Generating Function}\label{sec:theory}

We use the following notation: $\mathbb{N}$ denotes the positive
integers and $\mathbb{N}_{0}=\{0\}\cup\mathbb{N}$. We use $z=s+it\in\mathbb{C}$,
where $s=\operatorname{Re}(z)>0$. We use $\mathbb{E}[X^{k}]$ to denote integer moments, $k\in\mathbb{N}_{0}$, whereas $\mathbb{E}[X^{r}]$ denotes general moments including fractional and complex moments, $r\in\mathbb{C}$. For complex powers $z^{-(r+1)}$, we fix the principal branch of the complex logarithm (branch cut on $(-\infty,0]$) and
interpret integrals as oscillatory (i.e., symmetric Cauchy principal values) whenever $-1<\operatorname{Re}(r)\leq0$. Assumptions on one-sided exponential moments of $X$ (i.e., $\mathbb{E}[e^{\pm sX}]<\infty$ for some $s>0$) are stated with each theorem. When $\operatorname{Re}(r)\in(-1,0)$,
we additionally assume $\mathbb{E}\lvert X-\xi\rvert^{\operatorname{Re}(r)}<\infty$ to justify exchanging expectation and integration. 

Our results are based on the Fourier-Laplace transform of the law of $X$, $M_{X}(s+it)=\mathbb{E}[e^{(s+it)X}]\in\mathbb{C}$, that nests both the standard moment-generating function, $M_{X}(s)$, and the characteristic function, $\varphi_{X}(t)\equiv\mathbb{E}[e^{itX}]$, as special cases. 
To extract moments from this transform, we utilize the following global integral representations for various types of moments. 

\begin{lemma}\label{lem:identities}
For any 
$x\in\mathbb{R}\setminus\{0\}$ and any complex power $\operatorname{Re}(r)>-1$, 
we have the identity:
\begin{eqnarray}
    |x|^{r} &=& \frac{\Gamma(r+1)}{2\pi}\int_{-\infty}^{+\infty}\frac{e^{zx}+e^{-zx}}{z^{r+1}}\mathrm{d}t,\label{eq:abs_power}\\
    x_{+}^{r} &=& \frac{\Gamma(r+1)}{2\pi}\int_{-\infty}^{+\infty}\frac{e^{zx}}{z^{r+1}}\mathrm{d}t,\label{eq:pos_power}\\
    x^{k} &=& \frac{k!}{2\pi}\int_{-\infty}^{+\infty}\frac{e^{zx}+(-1)^{k}e^{-zx}}{z^{k+1}}\mathrm{d}t,\qquad k\in\mathbb{N}_{0}.\label{eq:signed_integer_power}
\end{eqnarray}
where $z=s+it$ with $s>0$ an arbitrary constant.

\end{lemma}
While the integrals in Lemma \ref{lem:identities} are absolutely convergent for $\operatorname{Re}(r)>0$, 
extending their validity to $-1 < \operatorname{Re}(r) \leq 0$ relies on evaluating the symmetric 
Cauchy principal value, $\lim_{T\to\infty}\int_{-T}^{T}$.\footnote{For $x=0$, the absolute-power identity holds for $\operatorname{Re}(r)>0$ and for $r=0$ under the convention $0^0=1$. The positive-part convention at the boundary is stated separately below.}

By taking the expectation of both sides of the identity in (\ref{eq:abs_power}), we obtain our first main result concerning fractional absolute moments. For $\operatorname{Re}(r)>0$, interchanging the expectation and integral is straightforward via Fubini's theorem. For $\operatorname{Re}(r)\in(-1,0]$, as noted, the improper integral must be interpreted as a symmetric Cauchy principal value, requiring $\mathbb{E}\lvert X-\xi\rvert^{\operatorname{Re}(r)}<\infty$ and $\Pr(X=\xi)=0$.

\begin{theorem}[Absolute moments]
\label{thm:factional-abs-moments}Suppose that $\mathbb{E}[e^{\pm sX}]<\infty$
for some $s>0$. For $\operatorname{Re}(r)\in(-1,0)$, assume additionally
that $\mathbb{E}|X-\xi|^{\operatorname{Re}(r)}<\infty$. If $\Pr(X=\xi)=0$ then for $r\in\mathbb{R}$ with $r>-1$
\begin{equation}
\mathbb{E}|X-\xi|^{r}
    = \tfrac{\Gamma(r+1)}{\pi}\int_{0}^{\infty}\operatorname{Re}\left[\frac{e^{-\xi z}M_{X}(z)+e^{\xi z}M_{X}(-z)}{z^{r+1}}\right]\mathrm{d}t,\qquad z=s+it,
    \label{eq:moment-r-abs-real}
\end{equation}
and for $r\in\mathbb{C}$ with $\operatorname{Re}(r)>-1$
\begin{equation}
\mathbb{E}|X-\xi|^{r}
    =\frac{\Gamma(r+1)}{2\pi}\int_{-\infty}^{+\infty}\frac{e^{-\xi z}M_{X}(z)+e^{\xi z}M_{X}(-z)}{z^{r+1}}\mathrm{d}t,\qquad z=s+it.
    \label{eq:moment-r-abs}
\end{equation}
Moreover, if $\Pr(X=\xi)>0$, the identities hold for $r\in\mathbb{R}$ with $r\geq0$ and for $r\in\mathbb{C}$ with $\operatorname{Re}(r)>0$, respectively.
\end{theorem}

For the case $r = 0$, the convention $0^{0} = 1$ yields $\mathbb{E}|X - \xi|^{0} = 1$, whereas for the limit $r \downarrow 0$, the right-limit can differ if the distribution contains an atom at $X = \xi$.

The finite moment requirement, $\mathbb{E}|X-\xi|^{\operatorname{Re}(r)}<\infty$, is implied by $\mathbb{E}[e^{\pm sX}]<\infty$ for $\operatorname{Re}(r)>0$, but is not guaranteed for $\operatorname{Re}(r)<0$. The results in Theorem \ref{thm:factional-abs-moments} include regular absolute moments by setting $\xi=0$ and central moments by setting $\xi=\mathbb{E}X$. Although analytical integration is often impractical, (\ref{eq:moment-r-abs}) offers a novel method to evaluate moments numerically. Numerical integration is, in our experience, insensitive to the choice of $s$, so long as $\mathbb{E}[e^{\pm sX}]<\infty$. 

For a random variable that is bounded from below, $\Pr(X\geq \xi)=1$, the absolute value operator is redundant. If $\Pr(X=\xi)=0$, the additional term in (\ref{eq:moment-r-abs}) vanishes by the support of the distribution, and the fractional moment simplifies to
$$
\mathbb{E}[(X-\xi)^r]
=
\frac{\Gamma(r+1)}{2\pi}
\int_{-\infty}^{+\infty}
\frac{e^{-\xi z}M_X(z)}{z^{r+1}}\mathrm{d}t,
\qquad \operatorname{Re}(r)>-1,
$$
whenever $\mathbb{E}[(X-\xi)^{\operatorname{Re}(r)}]<\infty$. If $\Pr(X=\xi)>0$, the formula remains valid for $\operatorname{Re}(r)>0$.\footnote{At $r=0$, a boundary atom contributes one half in the one-sided contour formula, so the formula is not valid under the convention $0^0=1$ unless $\Pr(X=\xi)=0$.}

A highly useful extension of Theorem \ref{thm:factional-abs-moments} arises in the context of exponential tilting (or the Esscher transform), which is central to changing probability measures in mathematical finance and extreme value theory.
By shifting the evaluation point of the MGF, we can extract exponentially tilted fractional moments directly.
\begin{corollary}[Exponentially Tilted Fractional Moments]
\label{cor:tilted_moments} Suppose that $\mathbb{E}[e^{(\pm s + v)X}]<\infty$ for some $s>0$ and a real tilting parameter $v$. Under the conditions of Theorem \ref{thm:factional-abs-moments}, the exponentially tilted fractional absolute moment, for $r\in\mathbb{C}$ with $\operatorname{Re}(r)>-1$, 
$$
\mathbb{E}\left[|X-\xi|^r e^{vX}\right] = \frac{\Gamma(r+1)}{2\pi}\int_{-\infty}^{+\infty}\frac{e^{-\xi z}M_{X}(z+v)+e^{\xi z}M_{X}(-z+v)}{z^{r+1}}\mathrm{d}t.
$$
\end{corollary}
For real $r$, this may equivalently be written as the folded real-part integral over $t\geq0$, as in Theorem \ref{thm:factional-abs-moments}.

For $v\in\mathbb{R}$ with $M_X(v)<\infty$, the Esscher-tilted measure is defined by $\mathrm{d}\mathbb{P}^{(v)}/\mathrm{d}\mathbb{P}=e^{vX}/M_X(v)$, so that $\mathbb{E}^{(v)}[g(X)]=\mathbb{E}[g(X)e^{vX}]/M_X(v)$.
If, in addition, $s>0$ is such that $M_X(v\pm s)<\infty$, then Corollary \ref{cor:tilted_moments} provides a contour integral representation for the numerator $\mathbb{E}[|X-\xi|^{r}e^{vX}]$, and hence for the tilted moment $\mathbb{E}^{(v)}[|X-\xi|^{r}]$.

The same contour bounds used above also justify differentiation with respect to the moment order $r$, provided the differentiated moment is finite. Since
$$
\frac{\partial}{\partial r}|X-\xi|^r = |X-\xi|^r\log|X-\xi|,
$$
differentiating the contour representation yields integral expressions for power-weighted logarithmic moments, with the derivative of $\Gamma(r+1)$
introducing the digamma function. The special case $r=0$ gives an expression for $\mathbb{E}[\log|X-\xi|]$.
\begin{corollary}[Logarithmic Moments]
\label{cor:log_moments} Suppose that $\mathbb{E}[e^{\pm sX}]<\infty$ for some $s>0$, and let
$r\in\mathbb{C}$ with $\operatorname{Re}(r)>-1$. If $\operatorname{Re}(r)\leq0$,
assume further that $\Pr(X=\xi)=0$ and
$$
\mathbb{E}\left[|X-\xi|^{\operatorname{Re}(r)}
\left(1+\left|\log|X-\xi|\right|\right)\right]<\infty.
$$
Then the fractional log-moment for $r\in\mathbb{C}$ with $\operatorname{Re}(r)>-1$ is given by,
\begin{equation}
\mathbb{E}\left[|X-\xi|^r \log|X-\xi|\right] = \psi(r+1)\mathbb{E}|X-\xi|^r - \tfrac{\Gamma(r+1)}{2\pi} \int_{-\infty}^{+\infty} \tfrac{e^{-\xi z}M_X(z) + e^{\xi z}M_X(-z)}{z^{r+1}} \log(z) \mathrm{d}t,
\end{equation}
where $\psi(\cdot)$ is the digamma function and where 
$\log(z)$ is the principal logarithm. For $r=0$, this yields the expression
\begin{equation}
\mathbb{E}[\log|X-\xi|] = -\gamma - \frac{1}{2\pi} \int_{-\infty}^{+\infty} \frac{e^{-\xi z}M_X(z) + e^{\xi z}M_X(-z)}{z} \log(z) \mathrm{d}t,
\end{equation}
where $\gamma$ is the Euler-Mascheroni constant.
\end{corollary}

The preceding results are based on the absolute-power contour identity. This identity is closely connected to the Parabolic Mellin Transform (PMT) developed in \citet{HansenTong:2026PMT}, who derived the globally valid integral representation
$$
\frac{1}{\Gamma(r+1)}
=
\frac{1}{\pi}
\int_{-\infty}^{+\infty}
\frac{e^{z^2}}{z^{2r+1}}\mathrm{d}t,
\qquad r\in\mathbb{C}.
$$
Combining this identity with Theorem \ref{thm:factional-abs-moments} yields a
Gamma-free representation of the fractional absolute moment:
$$
\mathbb{E}|X-\xi|^r
=
\frac{1}{2}
\frac{
\int_{-\infty}^{+\infty}
\frac{e^{-\xi z}M_X(z)+e^{\xi z}M_X(-z)}{z^{r+1}}\mathrm{d}t
}{
\int_{-\infty}^{+\infty}
\frac{e^{w^2}}{w^{2r+1}}\mathrm{d}\tau
},
$$
where the numerator is evaluated on $z=s+it$ and the denominator on any vertical line $w=\sigma+i\tau$, $\sigma>0$.
This Gamma-free form illustrates how the moment identities and the PMT
framework are two sides of the same contour-integral mechanism: PMT identities
for special functions can be used to normalize moment formulas, while moment
formulas provide probabilistic interpretations of the same contour integrals.

Below we will derive an expression for positive-part moments, where
we use the convention $x_{+}^{r}=x^{r}1_{\{x>0\}}$. This enables
us to accommodate negative powers, because $0_{+}^{r}=0$ for all
$r$.\footnote{Note that with this convention we have $0_{+}^{0}=0$, unlike $0^{0}=1$.}
Importantly, for $\operatorname{Re}(r)\in(-1,0]$, the integral $\int_{-\infty}^{+\infty}\frac{e^{-\xi z}M_{X}(z)}{z^{r+1}}\mathrm{d}t$
is to be interpreted as the symmetric Cauchy principal value, $\lim_{T\rightarrow\infty}\int_{-T}^{+T}\frac{e^{-\xi z}M_{X}(z)}{z^{r+1}}\mathrm{d}t$.

\begin{theorem}[Positive-part moments]
\label{thm:pos-part-moments}Suppose that $\mathbb{E}[e^{sX}]<\infty$
for some $s>0$. For $\operatorname{Re}(r)\in(-1,0)$, assume additionally
that $\mathbb{E}|X-\xi|^{\operatorname{Re}(r)}<\infty$. If $\Pr(X=\xi)=0$, then for $r\in\mathbb{C}$ with $\operatorname{Re}(r)>-1$,
\begin{equation}
\mathbb{E}[(X-\xi)_{+}^{r}]=\frac{\Gamma(r+1)}{2\pi}\int_{-\infty}^{+\infty}\frac{e^{-\xi z}M_{X}(z)}{z^{r+1}}\mathrm{d}t,\quad z=s+it.\label{eq:pos-var-moment}
\end{equation}
Moreover, if $\Pr(X=\xi)>0$ then the identity holds for $\operatorname{Re}(r)>0$.
\end{theorem}
Note that $\mathbb{E}[e^{sX}]<\infty$ for some $s>0$ ensures that $\mathbb{E}[(X-\xi)_{+}^{\operatorname{Re}(r)}]<\infty$ for ${\rm Re}\left(r\right)>0$ but not for $\operatorname{Re}(r)\in\left(-1,0\right)$. Therefore, for $\operatorname{Re}(r)\in\left(-1,0\right)$ where we rely on the dominated convergence theorem, the condition $\mathbb{E}|X-\xi|^{\operatorname{Re}(r)}<\infty$ is required for interchanging expectation and integral.

The expression in Theorem \ref{thm:pos-part-moments} is identical to that in \citet[theorem 1]{Pinelis:2011}, albeit we extend the range of moments to include some negative moments, $r\in(-1,0)$, and complex moments. The special case $r=1$ has received considerable attention because $\mathbb{E}[(X-\xi)_{+}]$ plays a central role in option pricing, with $\xi$ representing the strike price. For this special case, $r=1$, the real variant of (\ref{eq:pos-var-moment}) coincides with \citet[lemma 7.1]{KimRachevBianchiFabozzi:2010} and \citet[eq. 3.1]{HuangOosterlee:2011}.\footnote{These papers derived the result under slightly stronger assumptions. In \citet{KimRachevBianchiFabozzi:2010} it is assumed that the distribution is infinitely divisible and continuous, while \citet[eq. 3.1]{HuangOosterlee:2011} assume continuity of the distribution. Neither of these assumptions is required here.}

\begin{remark}
Here we assume $\mathbb{E}[e^{sX}]<\infty$ for some $s>0$, which does not guarantee the existence of the MGF in an open neighborhood of zero. So, $\mathbb{E}|X-\xi|^{\operatorname{Re}(r)}<\infty$ is not guaranteed by $\mathbb{E}[e^{sX}]<\infty$, as illustrated with the example below. Note that $\mathbb{E}[e^{sX}]<\infty$ for some $s>0$ does ensure $\mathbb{E}[(X-\xi)_{+}^{\operatorname{Re}(r)}]<\infty$ for $\operatorname{Re}(r)>0$.
\end{remark}

\begin{example}[One-sided MGF and divergent moments.]
Let $Y\sim\mathrm{Exp}(1)$, $Z\sim\mathrm{Pareto}(\alpha)$ with $\mathbb{P}(Z>z)=z^{-\alpha}$ for $z\geq1$, and choose $0<\alpha<1$. Define
$$
X=\begin{cases}
+Y, & \text{with prob. }1/2,\\
-Z, & \text{with prob. }1/2,
\end{cases}
$$
with $Y$, $Z$ independent. Then, for $0<s<1$, $\mathbb{E}[e^{sX}]=\tfrac{1}{2}\mathbb{E}[e^{sY}]+\tfrac{1}{2}\mathbb{E}[e^{-sZ}]=\tfrac{1}{2}(1-s)^{-1}+\tfrac{1}{2}\mathbb{E}[e^{-sZ}]<\infty$. However, for $\tilde{s}<0$ we have $\mathbb{E}[e^{\tilde{s}X}]\ge\tfrac{1}{2}\mathbb{E}[e^{|\tilde{s}|Z}]=\tfrac{1}{2}\int_{1}^{\infty}e^{|\tilde{s}|z}\alpha z^{-\alpha-1}\mathrm{d}z=\infty$. Hence the MGF is not finite on any open neighborhood of 0. For $r\in(-1,\alpha)$ we have $\mathbb{E}|X|^{r}=\tfrac{1}{2}\Gamma(r+1)+\tfrac{1}{2}\frac{\alpha}{\alpha-r}$, while for $r\geq\alpha$, $\mathbb{E}[|X|^{r}]\ge\tfrac{1}{2}\mathbb{E}[Z^{r}]=\infty$. We can nevertheless apply Theorem \ref{thm:pos-part-moments} for $r>-1$ because $\mathbb{E}(X)_{+}^{r}=\tfrac{1}{2}\Gamma(r+1)$.
\end{example}

For random variables whose support may include negative numbers, we have the following result for integer moments.

\begin{theorem}[Integer moments]
\label{thm:integer-moments}Suppose $\mathbb{E}[e^{\pm sX}]<\infty$
for some $s>0$. Then
\begin{equation}
\mathbb{E}[(X-\xi)^{k}] = \frac{k!}{\pi}\int_{0}^{\infty}\operatorname{Re}\left[\frac{e^{-\xi z}M_{X}(z)+(-1)^{k}e^{\xi z}M_{X}(-z)}{z^{k+1}}\right]\mathrm{d}t,\quad\text{for }k\in\mathbb{N}_{0}.\label{eq:moment-k}
\end{equation}
\end{theorem}
Because $k$ is real, we have expressed the integral in the form $\frac{k!}{\pi}\int_{0}^{\infty}\operatorname{Re}[\cdot]\mathrm{d}t$, which tends to be simpler to evaluate numerically than the equivalent expression using the form $\frac{k!}{2\pi}\int_{-\infty}^{+\infty}[\cdot]\mathrm{d}t$.

\begin{remark*}
In some situations, Theorems \ref{thm:factional-abs-moments}, \ref{thm:pos-part-moments}, and \ref{thm:integer-moments} provide different expressions for the same moments. For instance, (\ref{eq:moment-r-abs}) and the parabolic representation (\ref{eq:moment-parabolic}) must agree when $X-\xi>0$ almost surely, and for integer moments, these expressions must also agree with (\ref{eq:moment-k}). This is indeed the case, because the additional term in (\ref{eq:moment-r-abs}) and the additional term in (\ref{eq:moment-k}) are both zero for positive variables. This cancellation is a direct implication of the vanishing identity,
\begin{equation*}
    \int_{-\infty}^{+\infty} e^{-itx} (s+it)^{-(r+1)} \mathrm{d}t = 0,\quad \forall x>0,\ \forall s>0,\ \forall r\in\mathbb{C}\text{ with }\operatorname{Re}(r)>-1,
\end{equation*}
see Lemma \ref{lem:VanishingIdentity}. When this identity is used inside the moment formulas for $-1<\operatorname{Re}(r)\leq0$, the corresponding contour integrals are evaluated by symmetric truncation.
\end{remark*}

For strictly positive random variables, we can leverage the parabolic geometry of the PMT framework by \citet{HansenTong:2026PMT} to derive an integral representation. By evaluating the complex extension of the MGF along a parabolic contour, we induce Gaussian decay in the integrand, entirely bypassing the need for Cauchy principal value arguments.

\begin{theorem}[Parabolic Representation of Moments]
\label{thm:parabolic-positive} Let $X$ be a random variable such that $\Pr(X > \xi) = 1$, and suppose $\mathbb{E}[e^{s X}]<\infty$ for some $s>0$. For any $r\in\mathbb{C}$ such that
$\mathbb{E}(X-\xi)^{\operatorname{Re}(r)}<\infty$, and when
$\operatorname{Re}(r)=0$, additionally
$\mathbb{E}\left[1+\left|\log(X-\xi)\right|\right]<\infty$, we have:
\begin{equation}
\mathbb{E}[(X-\xi)^r] = \frac{\Gamma(r+1)}{\pi} \int_{-\infty}^{\infty} \frac{e^{-\xi z^2} M_X(z^2)}{z^{2r+1}} \mathrm{d}t,\label{eq:moment-parabolic}
\end{equation}
where $z = \sqrt{s}+it$. At the poles of $\Gamma(r+1)$,
$r=-1,-2,\ldots$, the right-hand side is interpreted in the limiting sense, equivalently by analytic continuation in $r$.
\end{theorem}

\begin{remark}
This formula is particularly useful because the parabolic contour introduces Gaussian damping for each strictly positive realization. Since $z=\sqrt{s}+it$, we have $\operatorname{Re}(z^2)=s-t^2$. Thus, for each fixed realization $y=X-\xi>0$, the kernel satisfies
$|e^{yz^2}|=e^{y(s-t^2)}$, which decays as $e^{-yt^2}$ as $|t|\to\infty$. After taking expectations, however, the averaged kernel $\mathbb{E}\left[e^{(X-\xi)(s-t^2)}\right]$
need not exhibit uniform Gaussian decay, because its tail behavior also depends on how much probability mass $X-\xi$ places near zero. The Mellin-Laplace argument in the proof shows that, for negative powers, this tail behavior is controlled by the corresponding negative moment $\mathbb{E}[(X-\xi)^{\operatorname{Re}(r)}]$. Thus the parabolic contour provides strong numerical damping for each positive realization, while the stated moment conditions control the boundary behavior near $X=\xi$.
\end{remark}

At the poles $r=-m$, $m=1,2,\ldots$, the limiting representation can be written directly as follows.
\begin{corollary}[Negative integer moments]
\label{cor:parabolic-negative-integers}
Let the assumptions of Theorem \ref{thm:parabolic-positive} hold, and write
$Y=X-\xi>0$. For $m=1,2,\ldots$, suppose additionally that
$\mathbb{E}[Y^{-m}(1+|\log Y|)]<\infty$. Then
$$
\mathbb{E}[(X-\xi)^{-m}]
=
\frac{2(-1)^m}{\pi(m-1)!}
\int_{-\infty}^{\infty}
\log(z)z^{2m-1}e^{-\xi z^2}M_X(z^2)\mathrm{d}t,
\qquad z=\sqrt{s}+it,
$$
where $\log(z)$ denotes the principal logarithm.
\end{corollary}

A numerical verification for negative moments of a chi-square distribution is provided in the Supplementary Material. The check includes both limiting evaluations of Theorem \ref{thm:parabolic-positive} near the poles of $\Gamma(r+1)$ and direct evaluations at the negative integers using the pole-free logarithmic contour formula in Corollary \ref{cor:parabolic-negative-integers}.

\section{Applications to Intractable Distributions}\label{sec:applications}

In this section, we illustrate the theoretical scope and computational advantages of the integral representations derived in Section \ref{sec:theory}. We apply the framework to two distributions widely used in applied probability and mathematical finance: the normal-inverse Gaussian (NIG) distribution and the compound Poisson-Gamma distribution. For both distributions, the moment-generating function is available in closed form, while traditional moment computation involves analytical complications. Fractional moments of the NIG distribution are typically expressed through infinite series of modified Bessel functions of the second kind. The compound Poisson-Gamma distribution is semi-continuous, with a point mass at zero and no closed-form density for the positive component, making density-based quadrature not directly applicable. We show that the CMGF framework handles both cases through the same contour-integral mechanism, yielding exact moments with substantial computational efficiency.

\subsection{Fractional Moments of the Normal-Inverse Gaussian Distribution}

To illustrate the method, we first use Theorem \ref{thm:factional-abs-moments} to compute fractional moments of the normal-inverse Gaussian (NIG) distribution. The NIG distribution, introduced in \citet{Barndorff-Nielsen:1978}, has four parameters: $\mu$ (location), $\delta$ (scale), $\alpha$ (tail heaviness), and $\beta$ (asymmetry). It has the density
$$
f(x)=\tfrac{\alpha\delta}{\pi\sqrt{\delta^{2}+(x-\mu)^{2}}}K_{1}\left(\alpha\sqrt{\delta^{2}+(x-\mu)^{2}}\right)e^{\delta\gamma+\beta(x-\mu)},\qquad x\in\mathbb{R},
$$
where $\gamma=\sqrt{\alpha^{2}-\beta^{2}}$ and $K_{1}(\cdot)$ is the modified Bessel function of the second kind. The corresponding MGF is given by $M_{X}(z)=\exp\left(\mu z+\delta\left[\gamma-\sqrt{\alpha^{2}-(\beta+z)^{2}}\right]\right)$, for $z=s+it$ with $-\alpha-\beta<s<\alpha-\beta$.

Existing expressions for the fractional moments of NIG distributions highlight the analytical bottlenecks of traditional approaches. Let $X\sim\mathrm{NIG}(\alpha,\beta,\mu,\delta)$ and $Y\sim\mathrm{NIG}(\alpha,0,\mu,\delta)$. \citet{Barndorff-NielsenStelzer:2005} showed that $\mathbb{E}|X|^{r}=e^{\delta(\gamma-\alpha)}\sum_{k=0}^{\infty}\frac{\beta^{k}}{k!}\mathbb{E}\left(|Y|^{r}(Y-\mu)^{k}\right)$ for $r>0$, expressing the moment as an infinite sum of moments involving the symmetric NIG distribution. For $\mu$-centered moments, \citet[corollary 3]{Barndorff-NielsenStelzer:2005} obtained the explicit formula:
\begin{equation}
\mathbb{E}(|X-\mu|^{r})=\frac{\alpha}{\pi}\left(\frac{2\delta}{\alpha}\right)^{\tfrac{r+1}{2}}e^{\delta\gamma}\sum_{k=0}^{\infty}\frac{2^{k}\Gamma(k+\tfrac{r+1}{2})}{(2k)!}\left(\frac{\delta\beta^{2}}{\alpha}\right)^{k}K_{k+\tfrac{r-1}{2}}(\delta\alpha),\quad r>0.\label{eq:InfSum-NIG-expression}
\end{equation}
Evaluating this infinite series of modified Bessel functions can be analytically cumbersome and numerically demanding, especially for fractional powers. By contrast, the CMGF integral reduces this entire series to a single, stable contour evaluation:
$$
\mathbb{E}|X-\mu|^{r}=\frac{\Gamma(r+1)}{2\pi}e^{\delta\gamma}\int_{-\infty}^{+\infty}\frac{e^{-\delta\sqrt{\alpha^{2}-(\beta-z)^{2}}}+e^{-\delta\sqrt{\alpha^{2}-(\beta+z)^{2}}}}{z^{r+1}}\mathrm{d}t,\qquad r>-1.
$$
A useful feature of the integral representation is that it accommodates arbitrary recentering, rather than only moments centered at $\mu$. This flexibility is not available in the infinite-sum expression in (\ref{eq:InfSum-NIG-expression}), whose algebraic form is tied to $\mu$-centered moments.

\begin{figure}[tbh]
\centering{}\includegraphics[width=1\textwidth]{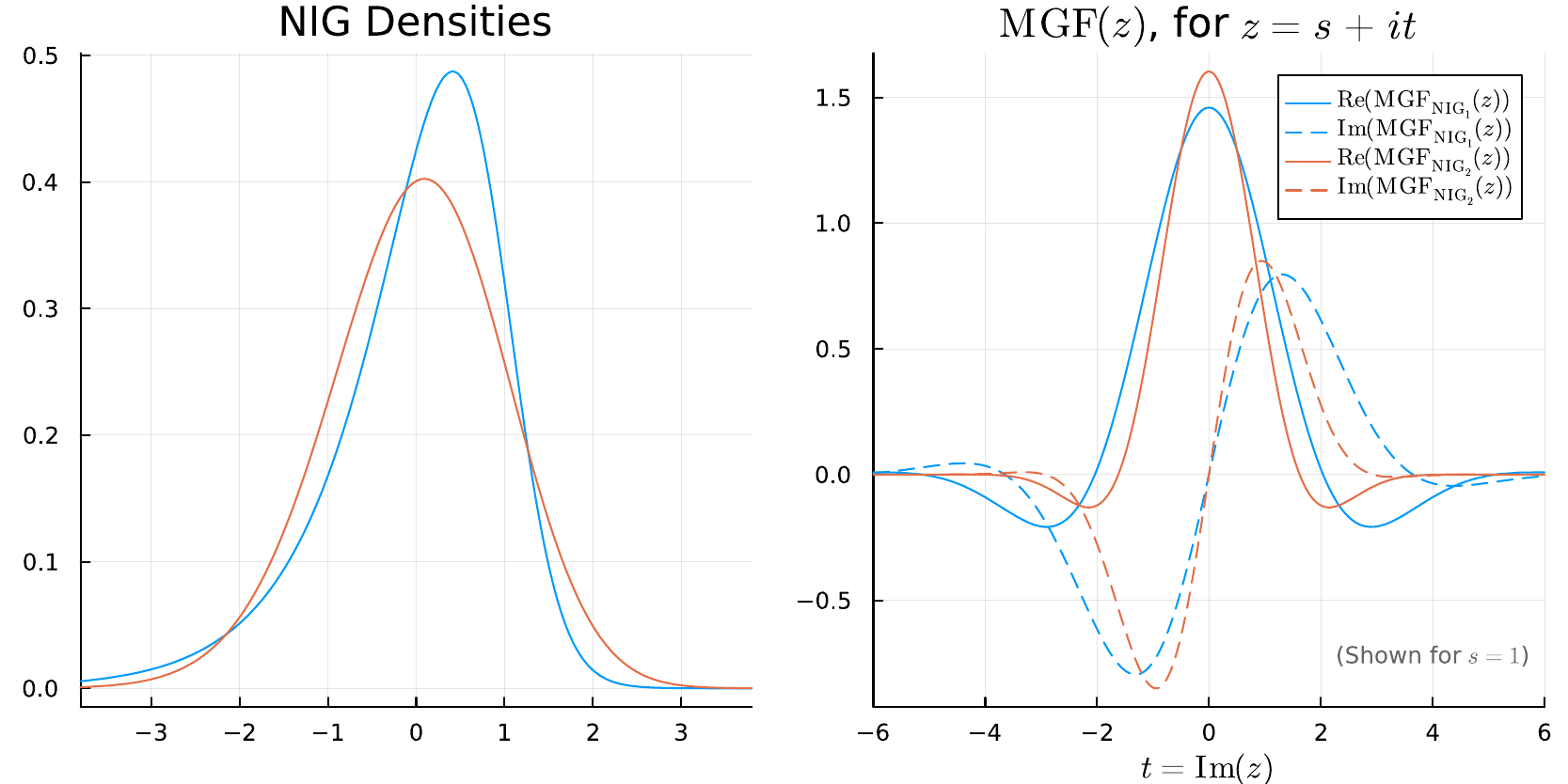}
\caption{Densities of two standardized NIG distributions (left panel) and the complex contour evaluation $t\mapsto M_X(1+it)$ (right panel). The two distributions correspond to shape parameters $(\xi,\chi)=(1/2,-1/3)$ (blue lines) and $(\xi,\chi)=(1/8,-1/16)$ (red lines).\label{fig:NIG_dens_MGF}}
\end{figure}

To illustrate the method's accuracy, we evaluate two standardized NIG distributions (zero mean and unit variance) characterized by the shape parameters $(\xi,\chi)=(1/2,-1/3)$ and $(\xi,\chi)=(1/8,-1/16)$ \citep{BNBJS1985nig}. Here we follow standard NIG notation; the NIG shape parameter $\xi$ should not be confused with the centering point $\xi$ used in Section 2. The exact algebraic mappings from $(\xi,\chi)$ to the standard parameters $(\alpha, \beta, \mu, \delta)$ are detailed in the Supplementary Material. 

The respective probability densities are shown in the left panel of Figure \ref{fig:NIG_dens_MGF}. To evaluate the CMGF integral, we set the real shift to $s=1$. For the NIG MGF, $M_X(z)=\exp\{\mu z+\delta(\gamma-\sqrt{\alpha^2-(\beta+z)^2})\}$, the strip of analyticity is $-\alpha-\beta<\operatorname{Re}(z)<\alpha-\beta$. Since Theorem \ref{thm:factional-abs-moments} involves both $M_X(z)$ and $M_X(-z)$, the contour shift must satisfy $0<s<\min\{\alpha-\beta,\alpha+\beta\}=\alpha-|\beta|$. For the two standardized distributions considered here, these upper limits are approximately $1.04$ and $5.29$, respectively, so $s=1$ is admissible in both cases. The right panel of Figure \ref{fig:NIG_dens_MGF} visualizes the complex MGF evaluated along this vertical contour, $t\mapsto M_X(1+it)$. The rapid decay and bounded oscillation of the real (solid) and imaginary (dashed) components provide a visual check of the integrability conditions underpinning Theorem \ref{thm:factional-abs-moments}.
\begin{figure}[tbh]
\begin{centering}
\includegraphics[width=1\textwidth]{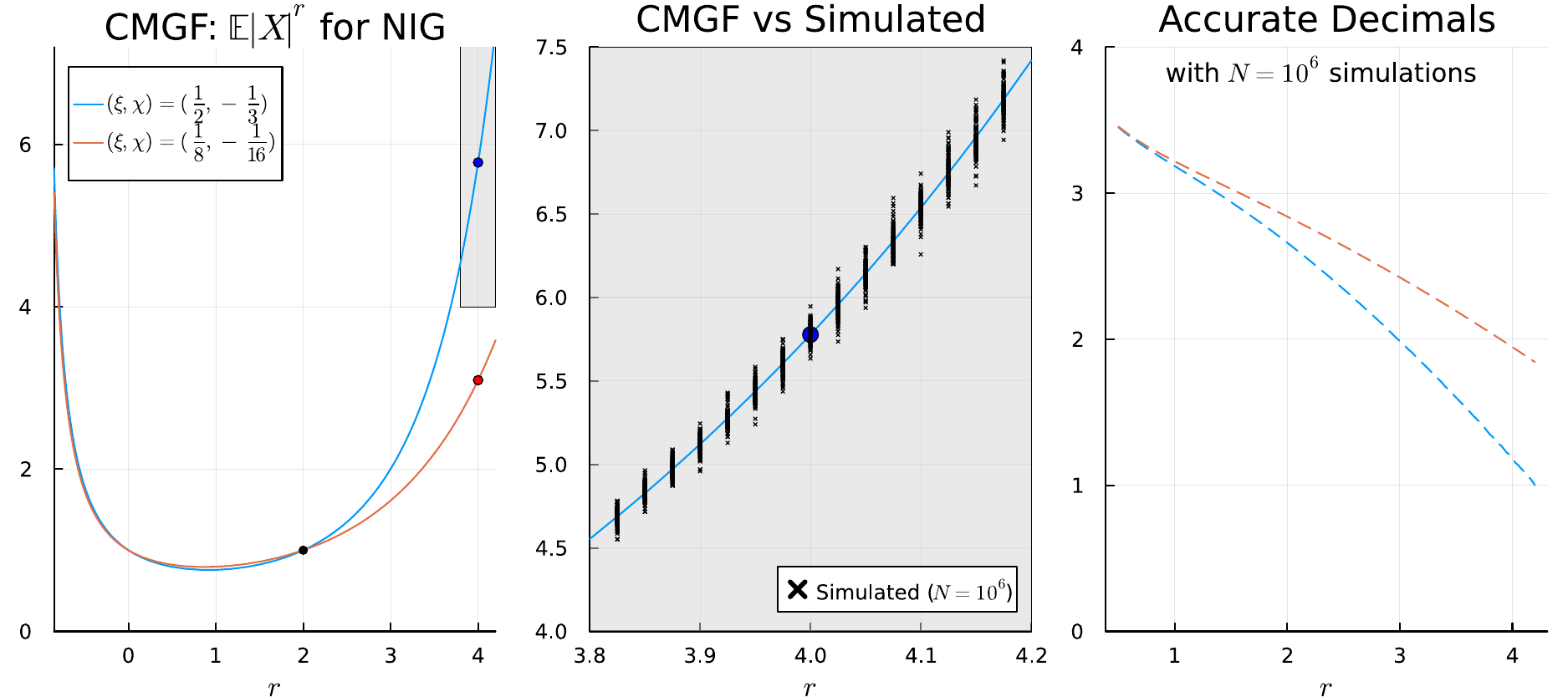}
\caption{Absolute moments of the standardized NIG distributions for $(\xi,\chi)=(1/2,-1/3)$ (blue lines) and $(\xi,\chi)=(1/8,-1/16)$ (red lines). The left panel shows the moments over $-0.85 \leq r \leq 4.2$, with dots representing the analytically known moments for $r=2$ and $r=4$. The middle panel overlays these curves with $\times$-crosses representing 100 simulation-based estimates of $\mathbb{E}|X|^{r}$ (each based on $N=1,000,000$ draws). The right panel shows the corresponding number of accurate decimal places for the simulation-based estimates.\label{fig:MomentsNIG}}
\par\end{centering}
\end{figure}

We compute the absolute moments for $-0.85\leq r\leq4.2$ using Theorem \ref{thm:factional-abs-moments}, as shown in the left panel of Figure \ref{fig:MomentsNIG}. Notably, the stable extraction of moments for $r\in(-1,0]$ provides a direct application of the symmetric Cauchy principal value and zero-mass conditions formalized in Section 2. The known second absolute moment (one) and fourth absolute moment, $3(1+4\chi^{2})/(1-\xi^{2})$, are recovered to near machine precision.

To assess the accuracy of simulation-based alternatives, we compare the CMGF moments with Monte Carlo estimates based on $N=1,000,000$ independent draws from the NIG distributions. The middle panel of Figure \ref{fig:MomentsNIG} overlays repeated simulation-based estimates on the CMGF curves, while the right panel reports the corresponding number of accurate decimal places. The simulation-based estimates deteriorate rapidly as $r$ increases, yielding only a few accurate decimal places even with a million draws.

The NIG example illustrates the main computational simplification of the CMGF representation: a fractional-moment problem that is traditionally expressed through an infinite series of modified Bessel functions is reduced to a single contour integral involving the closed-form MGF. The timing comparisons, Monte Carlo accuracy details, and implementation settings are reported in Section \ref{app:computational_details} of the Supplementary Material.

We next consider a compound distribution whose full law is mixed and whose positive-component density is not available in closed form. This creates a different obstacle for density-based methods and highlights the usefulness of the CMGF framework for semicontinuous distributions with a point mass at the boundary.

\subsection{Fractional Moments of the Compound Poisson-Gamma Distribution}

Next, we apply the integral framework to 
the classical compound Poisson-Gamma distribution, also known as the Tweedie distribution with index $1 < p < 2$ \citep{Tweedie:1984, Jorgensen:1987}. This model is widely used for modeling semi-continuous data because it naturally combines a point mass at zero with a continuous positive component. Let $N\sim\operatorname{Poisson}(\lambda)$ and $Y_{1},Y_{2},\ldots\sim\operatorname{Gamma}(\alpha,\beta)$ be i.i.d., and define the sum $X=\sum_{i=1}^{N}Y_{i}$ (with $X=0$ when $N=0$).

This distribution presents two obstacles for classical moment computation. First,
the law is mixed, with a point mass at zero and a continuous component on the
positive real line, so the full distribution does not admit an ordinary
Lebesgue density. Second, the density of the positive component is not available
in closed form. Density-based quadrature is therefore not directly applicable.
Although integer moments can be extracted recursively from cumulants,\footnote{For
instance, $\mathbb{E}(X^{4})=\kappa_{4}+4\kappa_{3}\kappa_{1}+3\kappa_{2}^{2}
+6\kappa_{2}\kappa_{1}^{2}+\kappa_{1}^{4}$, where
$\kappa_{k}=\lambda\mathbb{E}(Y^{k})$.} fractional moments are naturally
represented by conditioning on the positive component:
$$
\mathbb{E}[X^{r}|X>0] = \sum_{n=1}^{\infty}
\Pr(N=n|N>0)\beta^{r}\frac{\Gamma(n\alpha+r)}{\Gamma(n\alpha)}.
$$
This expression is exact, but it replaces one intractability by another: the
moment is represented by an infinite series whose terms involve Gamma-function
ratios. This representation is finite for $\operatorname{Re}(r)>-\alpha$, reflecting the behavior of the one-jump Gamma component near zero.

The CMGF approach avoids both obstacles. While the density is unavailable, the
moment-generating function of $X$ is explicit:
$M_{X}(z) = \exp\left(\lambda\left((1-\beta z)^{-\alpha}-1\right)\right)$, for $\operatorname{Re}(z)<1/\beta$. 
Moreover, the MGF of the conditional positive component is simply
$$
M_{X|X>0}(z)
=
\frac{M_X(z)-e^{-\lambda}}{1-e^{-\lambda}}.
$$
Thus the fractional moments of the positive component can be obtained directly
from the contour formulas by using $M_{X|X>0}(z)$, without evaluating the
density of the continuous component and without truncating an infinite series.

Importantly, this application explicitly illustrates the necessity of the boundary conditions formalized in Section 2. Because the distribution possesses a strictly positive mass at zero, $\Pr(X=0)=e^{-\lambda}>0$, the zero-mass condition required for negative powers in Theorem \ref{thm:factional-abs-moments} is violated for the unconditional distribution. Consequently, negative powers must be evaluated after conditioning on the positive component. The conditional distribution satisfies the required zero-mass constraint, and its MGF is $M_{X|X>0}(z)$. The existence of conditional negative moments is governed by the behavior of the positive component near zero; in this compound Poisson-Gamma case, the one-jump Gamma component implies existence for $r>-\alpha$.

For calibration and numerical illustration, we reparameterize into the standard Tweedie form $(\mu,\phi,p)$, where $\mu=\mathbb{E}X$ is the mean, $\phi>0$ is a dispersion parameter, and $p\in(1,2)$ is an index that controls the mean-variance relationship, $\operatorname{Var}(X)=\phi\mu^{p}$. The exact mapping is given by $(\mu,p,\phi)=(\lambda\alpha\beta,(\alpha+2)/(\alpha+1),\beta(\alpha+1)(\lambda\alpha\beta)^{-1/(\alpha+1)})$, such that $\lambda=\frac{\mu^{2-p}}{\phi(2-p)}$.

We present results for two distinct designs. In the first design, we set $(\mu,\phi,p)=(1,1,1.5)$, yielding a moderate Poisson rate and a modest point mass at zero ($\Pr(X=0)=e^{-2}\approx13.5\%$). This creates a visible semi-continuous component while maintaining a roughly standard continuous law on the positive axis. In the second design, $(\mu,\phi,p)=(1,2,1.3)$, the probability of zero is much larger, $\Pr(X=0)=e^{-1/1.4}\approx49.0\%$, and the positive component is more right-skewed. This design stresses the semi-continuous nature of the distribution.
\begin{figure}[htb]
\centering{}\includegraphics[width=0.5\textwidth]{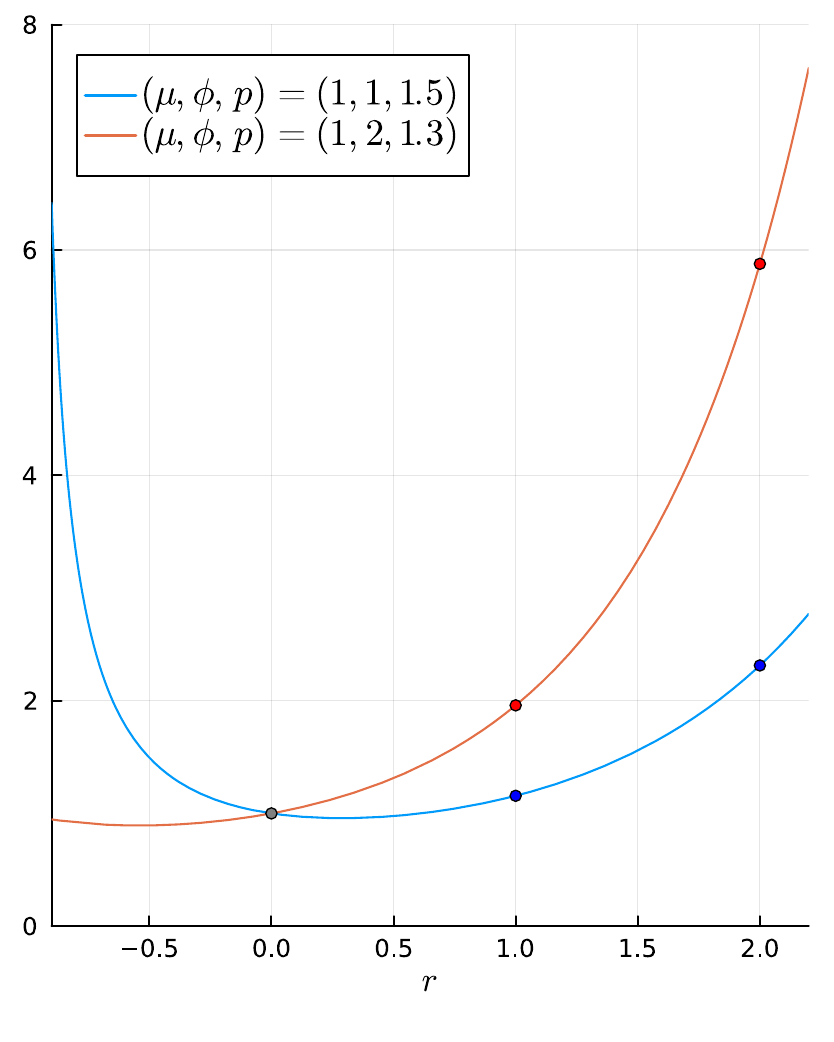}\includegraphics[width=0.5\textwidth]{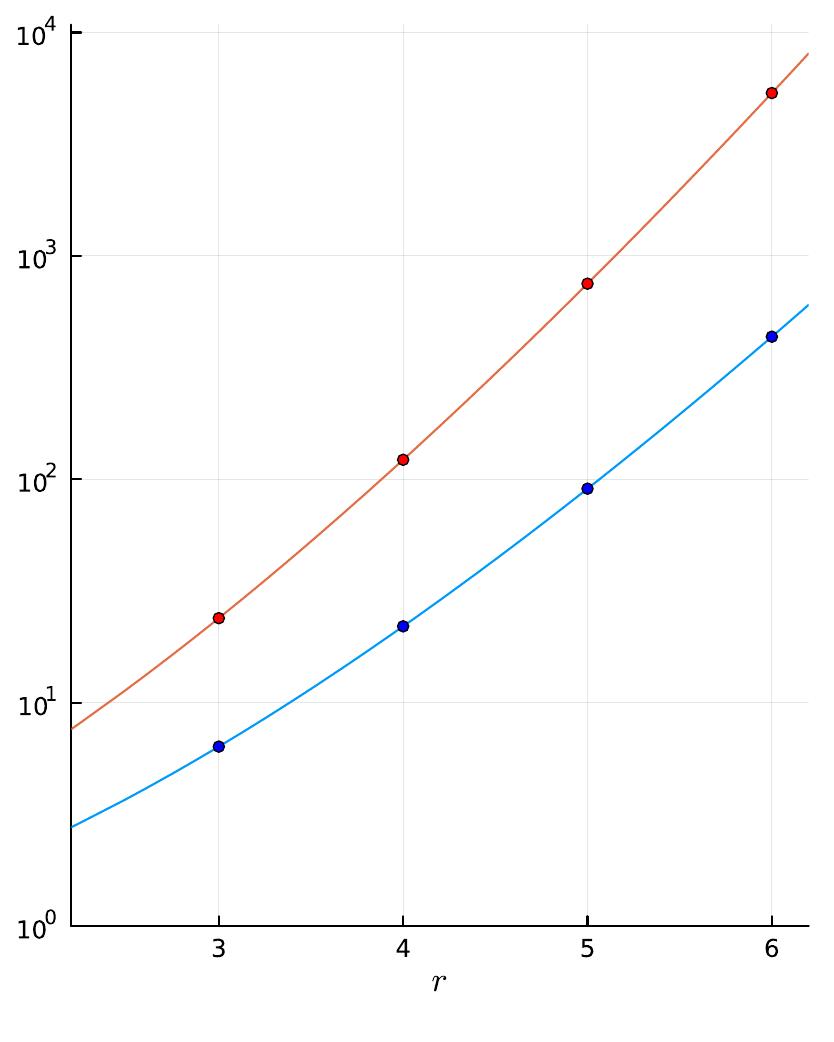}
\caption{{\small Fractional moments of two compound Poisson-Gamma distributions, parameterized by $(\mu,\phi,p)=(1,1,1.5)$ (blue) and $(\mu,\phi,p)=(1,2,1.3)$ (red). The conditional fractional moments, $\mathbb{E}[X^{r}|X>0]$, extracted via the CMGF framework are shown for negative and small positive $r$ in the left panel and for larger positive $r$ in the right panel (using a log-scale $y$-axis). The overlaid dots are exact conditional integer moments. At $k=0$, $\mathbb{E}[X^{0}|X>0]=1$; for $k=1,2,\ldots$, $\mathbb{E}[X^{k}|X>0]=\mathbb{E}[X^{k}]/(1-e^{-\lambda})$, where the unconditional integer moments $\mathbb{E}[X^{k}]$ are computed directly from exact cumulant expressions. These dots confirm the numerical precision of the CMGF integral method.\label{fig:MomentsCompound}}}
\end{figure}

As shown in Figure \ref{fig:MomentsCompound}, the CMGF method extracts the continuum of conditional fractional moments for both designs and intersects the known cumulant-derived integer moments. The left panel also illustrates the behavior of negative moments. In the first design, $p=1.5$ implies $\alpha=1$, so negative moments exist only for $r>-1$ and the curve rises sharply as $r$ approaches this boundary. In the second design, $p=1.3$ implies $\alpha=7/3$, so the existence boundary is farther left, at $r=-7/3$, and the negative-moment curve is correspondingly more stable near $r=0$. Thus the example illustrates both the atom-at-zero issue and the role of the positive-component boundary behavior in determining which negative moments exist. The contour representation avoids density evaluation and avoids truncating the infinite conditional series.

\section{Conclusion}
In this paper, we introduced a general integral framework for computing fractional, complex, negative, and logarithmic moments from the moment-generating function under explicit regularity conditions. By evaluating a complex extension of the MGF along vertical and parabolic contours, we derived a suite of exact identities that entirely bypass the need for explicit probability densities or analytically prohibitive successive derivatives. Crucially, we formalized the topological constraints governing these representations, establishing the necessity of the symmetric Cauchy principal value and the zero-mass condition for extracting negative fractional moments. Furthermore, we demonstrated the theoretical flexibility of this framework by extending it to extract logarithmic moments via the digamma function, defining moments purely as a ratio of contour integrals, and establishing stable representations with Gaussian decay for strictly positive variables.

The applications in Section 3 show that the framework turns analytically difficult moment calculations into tractable contour integrals involving the MGF, including settings where existing formulas are cumbersome or density-based methods are not directly available. For the normal-inverse Gaussian distribution, the complex contour integral replaces an infinite series of modified Bessel functions of the second kind with a single numerically stable evaluation. For the compound Poisson-Gamma (Tweedie) distribution, the method produces a continuum of fractional moments despite the absence of a closed-form density and the presence of a point mass at zero. In both cases, the mathematical simplification provided by the framework translates directly into substantial computational efficiency.

Future research could explore the extension of these contour integral representations to multivariate domains. While simple bivariate cross-moments can be extracted through univariate projections, generalizing the complex MGF framework to evaluate arbitrary joint fractional moments in multivariate settings represents a natural avenue for advancing the methods developed in this paper.

\bibliographystyle{apalike}
\bibliography{prh}

\clearpage

\appendix

\setcounter{equation}{0}
\global\long\def\theequation{A.\arabic{equation}}%

\setcounter{lemma}{0}
\global\long\def\thelemma{A.\arabic{lemma}}%

\setcounter{figure}{0}
\global\long\def\thefigure{A.\arabic{figure}}%


\section{Appendix of Proofs}

\begin{lemma}[Vanishing Identity]
\label{lem:VanishingIdentity}
Let $w = \sigma + it$ with $\sigma > 0$. For any $x > 0$ and any $r \in \mathbb{C}$ with $\operatorname{Re}(r) > -1$, we have:
\begin{equation*}
\int_{-\infty}^{\infty} \frac{e^{-itx}}{w^{r+1}} dt = 0.
\end{equation*}
\end{lemma}
\begin{proof}Set $a=\sigma$, $b=x$, and $\nu=r+1$ in the last identity of \citet[8.315.2]{GradshteynRyzhik:2007}, which reads
$\int_{-\infty}^{\infty}
\frac{e^{-ibt}}{(a+it)^\nu}\mathrm{d}t=0$,
for $\operatorname{Re}(a)>0$, $b>0$, and  $\operatorname{Re}(\nu)>0$.

A detailed proof is presented in Section \ref{sec:AuxContourBounds} of the Supplementary Material.
\end{proof}

\noindent \textbf{Proof of Lemma \ref{lem:identities}.}
We first prove the positive-part identity. Starting from the reciprocal-gamma Bromwich-Laplace representation \citep[see, e.g.,][]{Laplace:1812,Pribitkin:2002},
$$
\frac{1}{\Gamma(r+1)}
= \frac{1}{2\pi}\int_{-\infty}^{\infty}\frac{e^w}{w^{r+1}}\mathrm{d}\tau,\qquad w=a+i\tau,\quad a>0,
$$
valid for $\operatorname{Re}(r)>-1$, with the principal branch of the logarithm.

Let $x>0$ and set $w=zx$, where $z=s+it$. Then $w=sx+ixt$, so $\mathrm{d}\tau=x\mathrm{d}t$, and hence
$$
x^r = \frac{\Gamma(r+1)}{2\pi}
\int_{-\infty}^{\infty}
\frac{e^{zx}}{z^{r+1}}\mathrm{d}t.
$$
If $x<0$, then Lemma \ref{lem:VanishingIdentity} gives
$\int_{-\infty}^{\infty}\frac{e^{zx}}{z^{r+1}}\mathrm{d}t=0$.
Thus, for $x\neq0$,
$$
x_+^r
=
\frac{\Gamma(r+1)}{2\pi}
\int_{-\infty}^{\infty}
\frac{e^{zx}}{z^{r+1}}\mathrm{d}t.
$$

Applying the same identity to $-x$ gives
$$
(-x)_+^r
=
\frac{\Gamma(r+1)}{2\pi}
\int_{-\infty}^{\infty}
\frac{e^{-zx}}{z^{r+1}}\mathrm{d}t.
$$
Since $|x|^r=x_+^r+(-x)_+^r$ for $x\neq0$, it follows that
$$
|x|^r
=
\frac{\Gamma(r+1)}{2\pi}
\int_{-\infty}^{\infty}
\frac{e^{zx}+e^{-zx}}{z^{r+1}}\mathrm{d}t.
$$

Finally, let $k\in\mathbb{N}_0$. Applying the positive-part identity with $r=k$ to $x$ and $-x$ gives
$$
x_+^k
=
\frac{k!}{2\pi}
\int_{-\infty}^{\infty}
\frac{e^{zx}}{z^{k+1}}\mathrm{d}t,
\qquad
(-x)_+^k
=
\frac{k!}{2\pi}
\int_{-\infty}^{\infty}
\frac{e^{-zx}}{z^{k+1}}\mathrm{d}t.
$$
Since $x^k=x_+^k+(-1)^k(-x)_+^k$, we obtain
$$
x^k
=
\frac{k!}{2\pi}
\int_{-\infty}^{\infty}
\frac{e^{zx}+(-1)^k e^{-zx}}{z^{k+1}}\mathrm{d}t.
$$
For $\operatorname{Re}(r)>0$, the integrals are absolutely convergent. For $-1<\operatorname{Re}(r)\leq0$, they are interpreted as symmetric Cauchy principal values. This completes the proof.
\hfill{}$\square$\medskip{}

\noindent \textbf{Proof of Theorem \ref{thm:factional-abs-moments}.}
Setting $x=X-\xi$ in (\ref{eq:abs_power}) and taking expectations gives the formal identity. The proof below justifies the interchange of expectation and integration across the relevant domains of $r$.

\noindent \textit{Case $\operatorname{Re}(r)>0$:} We apply Tonelli's theorem to the absolute integral. For any $x\in\mathbb{R}$, the integrand is bounded using $\left|z^{-(r+1)}\right| \leq e^{\pi|\operatorname{Im}(r)|} |z|^{-(\operatorname{Re}(r)+1)}$. Thus,
$$
\int_{-\infty}^{+\infty}\mathbb{E}\left|\frac{e^{z(X-\xi)}+e^{-z(X-\xi)}}{z^{r+1}}\right|\mathrm{d}t \leq e^{\pi|\operatorname{Im}(r)|} \left[e^{-s\xi}\mathbb{E}[e^{sX}] + e^{s\xi}\mathbb{E}[e^{-sX}]\right] \int_{-\infty}^{+\infty}\frac{1}{(s^{2}+t^{2})^{(\operatorname{Re}(r)+1)/2}}\mathrm{d}t.
$$
The condition $\mathbb{E}[e^{\pm sX}]<\infty$ ensures the expectations are finite, and the deterministic integral is finite for $\operatorname{Re}(r)>0$. By Fubini's theorem, the interchange is justified. Note that if $\Pr(X=\xi)>0$, this domain remains valid because $0^r = 0$ for $\operatorname{Re}(r)>0$.

\noindent \textit{Case $\operatorname{Re}(r)\in(-1,0)$:}
Let $p=\operatorname{Re}(r)$. Because the integral is interpreted as a symmetric Cauchy principal value, define
$$
A_T(x;r)=
\int_{-T}^{T}
\frac{e^{zx}+e^{-zx}}{z^{r+1}}\mathrm{d}t.
$$
By Lemma \ref{lem:osc-bound} of the Supplementary Material, for $x\neq0$,
$$
\sup_{T>0}|A_T(x;r)|
\leq
C_r(e^{sx}+e^{-sx})|x|^p.
$$
With $x=X-\xi$, the right-hand side is integrable. Indeed, on $|X-\xi|\geq1$, $|X-\xi|^p\leq1$, while on $|X-\xi|<1$ the factors $e^{\pm s(X-\xi)}$ are bounded and integrability follows from the assumed negative moment $\mathbb{E}|X-\xi|^p<\infty$. Since $\Pr(X=\xi)=0$, the singular point is avoided almost surely. The dominated convergence theorem therefore permits the interchange of the principal-value limit and expectation.

\noindent \textit{Case $r=0$:}
For the one-sided truncated kernel,
$$
I_T(x)=\int_{-T}^{T}\frac{e^{(s+it)x}}{s+it}\mathrm{d}t.
$$
If $x=0$, explicit integration gives $I_T(0)=2\arctan(T/s)$, so $|I_T(0)|\leq\pi$. If $x\neq0$, set $u=t|x|$, $S=T|x|$, and $a=s|x|$. Then
$$
I_T(x)
=
e^{sx}
\int_{-S}^{S}
\frac{e^{iu\operatorname{sgn}(x)}}{a+iu}\mathrm{d}u.
$$
The scaled integral is uniformly bounded in $S>0$ and $a>0$. To see this, decompose it into cosine and sine parts:
$$
\int_{-S}^{S}
\frac{e^{\pm iu}}{a+iu}\mathrm{d}u
=
2\int_0^S
\frac{a\cos u\pm u\sin u}{a^2+u^2}\mathrm{d}u.
$$
The cosine term is bounded by $\arctan(S/a)\leq\pi/2$. For the sine term, use
$$
\frac{u}{a^2+u^2}=\frac{1}{u}-\frac{a^2}{u(a^2+u^2)}
$$
to obtain
$$
\left|\int_0^S\frac{u\sin u}{a^2+u^2}\mathrm{d}u\right|
\leq
\left|\int_0^S\frac{\sin u}{u}\mathrm{d}u\right|
+
\left|\int_0^S\frac{a^2\sin u}{u(a^2+u^2)}\mathrm{d}u\right|.
$$
The first term is bounded by $\operatorname{Si}(\pi)$. For the second term, the function $a^2/\{u(a^2+u^2)\}$ is positive and decreasing on $(0,\infty)$. Dirichlet's test for oscillatory integrals gives 
$$ \sup_{S>0} \left| \int_0^S \frac{a^2\sin u}{u(a^2+u^2)}\mathrm{d}u \right| \leq C $$ 
with a constant independent of $a$ and $S$. In particular, the full improper integral is $$ \int_0^\infty \frac{a^2\sin u}{u(a^2+u^2)}\mathrm{d}u = \frac{\pi}{2}(1-e^{-a}), $$ which is uniformly bounded in $a>0$.
Thus
$\sup_{T>0}|I_T(x)|\leq C e^{sx}$. 
Applying the same bound to the term with $e^{-zx}$ gives
$$
\sup_{T>0}
\left|
\int_{-T}^{T}
\frac{e^{zx}+e^{-zx}}{z}\mathrm{d}t
\right|
\leq
C(e^{sx}+e^{-sx}).
$$
With $x=X-\xi$, this dominating function is integrable under $\mathbb{E}[e^{\pm sX}]<\infty$. Hence dominated convergence justifies the interchange at $r=0$, including the case $\Pr(X=\xi)>0$, and the formula gives $\mathbb{E}|X-\xi|^0=1$ under the convention $0^0=1$.

\noindent \textit{Case $\operatorname{Re}(r)=0$, $\operatorname{Im}(r)\neq0$:}
Write $r=i\eta$ with $\eta\neq0$. We require $\Pr(X=\xi)=0$, since $0^{i\eta}$ is not defined. For $x\neq0$, set $u=t|x|$, $S=T|x|$, and $a=s|x|$. Then
$$
\int_{-T}^{T}
\frac{e^{(s+it)x}}{(s+it)^{1+i\eta}}\mathrm{d}t
=
e^{sx}|x|^{i\eta}
\int_{-S}^{S}
\frac{e^{iu\operatorname{sgn}(x)}}{(a+iu)^{1+i\eta}}\mathrm{d}u.
$$
The required uniform bound for the scaled integral is precisely the content of Lemma \ref{lem:boundary-osc-bound} of the Supplementary Material. Hence
$$
\sup_{T>0}
\left|
\int_{-T}^{T}
\frac{e^{zx}+e^{-zx}}{z^{1+i\eta}}\mathrm{d}t
\right|
\leq
C_\eta(e^{sx}+e^{-sx}).
$$
With $x=X-\xi$, this dominating function is integrable under $\mathbb{E}[e^{\pm sX}]<\infty$. Since $\Pr(X=\xi)=0$, dominated convergence justifies the interchange.

Combining the preceding cases yields
$$
\mathbb{E}|X-\xi|^r
=
\frac{\Gamma(r+1)}{2\pi}
\int_{-\infty}^{+\infty}
\frac{e^{-\xi z}M_X(z)+e^{\xi z}M_X(-z)}{z^{r+1}}\mathrm{d}t.
$$
For real $r$, the integrand at $-t$ is the complex conjugate of the integrand at $t$, since $X$ is real-valued and $M_X(\overline{z})=\overline{M_X(z)}$. Therefore the full-line integral folds to
$$
\mathbb{E}|X-\xi|^r
=
\frac{\Gamma(r+1)}{\pi}
\int_0^\infty
\operatorname{Re}
\left[
\frac{e^{-\xi z}M_X(z)+e^{\xi z}M_X(-z)}{z^{r+1}}
\right]\mathrm{d}t.
$$
This completes the proof.
\hfill{}$\square$\medskip{}

\noindent \textbf{Proof of Corollary \ref{cor:tilted_moments}.} The result follows immediately from Theorem \ref{thm:factional-abs-moments} by replacing the standard expectation operator with the exponentially tilted measure. The term $e^{zX}$ in the derivation merges with the tilting factor $e^{vX}$, mapping the required moment-generating function evaluation from $M_X(z)$ to $\mathbb{E}[e^{(z+v)X}] = M_X(z+v)$. The uniform convergence is preserved provided the shifted arguments $\pm s + v$ remain within the domain of convergence for the MGF.
\hfill{}$\square$\medskip{}

\noindent \textbf{Proof of Corollary \ref{cor:log_moments}.} 
Because differentiating the contour representation introduces an additional logarithmic factor, we first justify the differentiation under the integral sign. For $\operatorname{Re}(r)>0$, this follows from absolute convergence. Indeed, for $z=s+it$,
$$
\left|\log(z)\frac{e^{-\xi z}M_X(z)+e^{\xi z}M_X(-z)}{z^{r+1}}\right|\leq C_r\frac{1+\log(1+|t|)}{(s^2+t^2)^{(\operatorname{Re}(r)+1)/2}},
$$
where $C_r$ depends on $s$, $\xi$, $M_X(s)$, and $M_X(-s)$, but not on $t$. The right-hand side is integrable over $\mathbb{R}$ whenever $\operatorname{Re}(r)>0$.

For $\operatorname{Re}(r)\leq0$, the contour integrals are interpreted as symmetric Cauchy principal values, and the preceding absolute-integrability argument is not available without imposing additional decay assumptions on $M_X(s+it)$. We do not impose such assumptions. Instead, we differentiate the deterministic contour identity for $|x|^r$ with respect to $r$ and then apply it pointwise at $x=X-\xi$. The differentiated identity introduces the factor $\log(z)$ on the contour side and the factor $\log|x|$ on the power side. 
The logarithmic version of the oscillatory bound is proved in Lemma \ref{lem:log-osc-bound} of the Supplementary Material. It gives the truncated principal-value bound
$$
\sup_{T>0}\left|\int_{-T}^{T}\frac{e^{zx}+e^{-zx}}{z^{r+1}}\log(z)\mathrm{d}t\right|\leq C_r\left(e^{sx}+e^{-sx}\right)|x|^{\operatorname{Re}(r)}\left(1+\left|\log|x|\right|\right),
$$
for $x\neq0$. Hence, with $x=X-\xi$, the assumed integrability condition
$$
\mathbb{E}\left[|X-\xi|^{\operatorname{Re}(r)}\left(1+\left|\log|X-\xi|\right|\right)\right]<\infty
$$
and the exponential-moment assumptions imply that the dominating function is integrable. Dominated convergence therefore justifies the differentiation and the passage of the limit through the expectation. Thus no absolute continuity of the distribution, nor any explicit decay condition on $M_X(s+it)$, is required.

Differentiating both sides of (\ref{eq:moment-r-abs}) with respect to $r$ yields $\mathbb{E}[|X-\xi|^r \log|X-\xi|]$ on the left-hand side. Applying the product rule to the right-hand side using $\frac{\partial}{\partial r} \Gamma(r+1) = \Gamma(r+1)\psi(r+1)$ and $\frac{\partial}{\partial r} z^{-(r+1)} = -z^{-(r+1)}\log(z)$ yields the first identity. Substituting $r=0$ and using $\Gamma(1)=1$, $\mathbb{E}|X-\xi|^0=1$, and $\psi(1)=-\gamma$ simplifies directly to the second identity. \hfill $\square$

\noindent \textbf{Proof of Theorem \ref{thm:pos-part-moments}.}
We use the convention
$x_{+}^{r}\equiv x^{r}\mathds{1}_{\{x>0\}}$, so that $0_{+}^{r}=0$ for all $r$. By the positive-part identity in Lemma \ref{lem:identities}, for $x\neq0$ and $\operatorname{Re}(r)>-1$,
$$
x_{+}^{r}
=
\frac{\Gamma(r+1)}{2\pi}
\int_{-\infty}^{+\infty}
\frac{e^{zx}}{z^{r+1}}\mathrm{d}t,
\qquad z=s+it,\quad s>0.
$$
This identity is based on the reciprocal-gamma contour representation together with the vanishing identity in Lemma \ref{lem:VanishingIdentity}. At the boundary $x=0$, the identity extends only for $\operatorname{Re}(r)>0$, consistent with $0_{+}^{r}=0$. For $\operatorname{Re}(r)\leq0$, the point $x=0$ must be excluded.

Substituting $x=X-\xi$ and taking expectations, we justify the interchange of expectation and integration as follows.

\noindent \textit{Case $\operatorname{Re}(r)>0$}: We apply Tonelli's theorem to the absolute integral. Bounding the complex power as in Theorem \ref{thm:factional-abs-moments}, we have:
$$
\mathbb{E}\left[\int_{-\infty}^{+\infty}\left|\frac{e^{z(X-\xi)}}{z^{r+1}}\right|\mathrm{d}t\right]\leq e^{\pi|\operatorname{Im}(r)|}e^{-s\xi}\mathbb{E}\left[e^{sX}\right]\frac{\sqrt{\pi}\Gamma\left(\frac{\operatorname{Re}(r)}{2}\right)}{s^{\operatorname{Re}(r)}\Gamma\left(\frac{\operatorname{Re}(r)+1}{2}\right)}<\infty.
$$
Since $\mathbb{E}[e^{sX}]<\infty$, the bound is finite. Fubini's theorem therefore permits the interchange of expectation and integration, proving the result for $\operatorname{Re}(r)>0$. Notably, a point mass at the boundary, $\Pr(X=\xi)>0$, is permitted here because $0_+^r = 0$ is well-behaved for $\operatorname{Re}(r)>0$.

\noindent \textit{Case $\operatorname{Re}(r)\in(-1,0]$}: The integral is interpreted as a symmetric Cauchy principal value. For $p=\operatorname{Re}(r)\in(-1,0)$, the one-sided version of Lemma \ref{lem:osc-bound} gives $$ \sup_{T>0}\left|\int_{-T}^{T}\frac{e^{zx}}{z^{r+1}}\mathrm{d}t\right| \leq C_r e^{sx}|x|^p,\qquad x\neq0. $$ For the boundary case $p=0$, the corresponding estimate follows from Lemma \ref{lem:boundary-osc-bound}. Hence, with $x=X-\xi$, the truncated integrals are dominated by an integrable function, using $\mathbb{E}[e^{sX}]<\infty$ and, when $p<0$, the additional condition $\mathbb{E}|X-\xi|^p<\infty$. Because the identity at $x=0$ is not valid in this range under the positive-part convention, we assume $\Pr(X=\xi)=0$. The dominated convergence theorem then justifies the interchange of the principal-value limit and expectation.
\hfill{}$\square$\medskip{}

\noindent{}\textbf{Proof of Theorem \ref{thm:integer-moments}.}
We use the signed-power identity in (\ref{eq:signed_integer_power}). For $x\in\mathbb{R}$,
$k\in\mathbb{N}_{0}$, and $z=s+it$ with $s>0$,
$$
x^{k}
=
\frac{k!}{2\pi}
\int_{-\infty}^{+\infty}
\frac{e^{zx}+(-1)^k e^{-zx}}{z^{k+1}}
\mathrm{d}t.
$$
For $k\geq1$, the integral is absolutely convergent. For $k=0$, it is interpreted as the symmetric Cauchy principal value, consistently with the convention $x^0=1$.

Setting $x=X-\xi$ and taking expectations gives
\begin{align*}
\mathbb{E}[(X-\xi)^k]
&=
\frac{k!}{2\pi}
\int_{-\infty}^{+\infty}
\frac{
\mathbb{E}\left[e^{z(X-\xi)}\right]
+(-1)^k\mathbb{E}\left[e^{-z(X-\xi)}\right]
}{z^{k+1}}
\mathrm{d}t \\
&=
\frac{k!}{2\pi}
\int_{-\infty}^{+\infty}
\frac{
e^{-\xi z}M_X(z)
+(-1)^k e^{\xi z}M_X(-z)
}{z^{k+1}}
\mathrm{d}t.
\end{align*}
The interchange of expectation and integration is justified by the assumption $\mathbb{E}[e^{\pm sX}]<\infty$; for $k\geq1$ this follows from absolute convergence, and for $k=0$ from the same symmetric principal-value argument used in the proof of Theorem \ref{thm:factional-abs-moments}.

It remains to fold the integral to the positive half-line. Since $X$ is real-valued, $M_X(\overline{z})=\overline{M_X(z)}$.
Thus, along the vertical contour $z=s+it$, the integrand evaluated at $-t$ is the complex conjugate of the integrand evaluated at $t$. Therefore,
\begin{equation*}
\int_{-\infty}^{+\infty}
\frac{
e^{-\xi z}M_X(z)
+(-1)^k e^{\xi z}M_X(-z)
}{z^{k+1}}
\mathrm{d}t
=
2\int_0^\infty
\operatorname{Re}\left[
\frac{
e^{-\xi z}M_X(z)
+(-1)^k e^{\xi z}M_X(-z)
}{z^{k+1}}
\right]
\mathrm{d}t.
\end{equation*}
Substituting this into the preceding expression yields the stated formula.
\hfill{}$\square$\medskip{}

\noindent \textbf{Proof of Theorem \ref{thm:parabolic-positive}.} From \citet[Theorem 1]{HansenTong:2026PMT} we have for any $\rho\in\mathbb{C}$ and any $\sigma>0$, $$ \frac{1}{\Gamma(\rho)} = \frac{1}{\pi} \int_{-\infty}^{\infty} w^{1-2\rho}e^{w^2}\mathrm{d}\tau, \qquad w=\sigma+i\tau, $$ where the principal branch is used. Taking $\rho=r+1$ gives $$ \frac{1}{\Gamma(r+1)} = \frac{1}{\pi} \int_{-\infty}^{\infty} w^{-(2r+1)}e^{w^2}\mathrm{d}\tau, \qquad w=\sigma+i\tau,\quad \sigma>0. $$ 
Consider a fixed realization $y=X-\xi>0$. Since the reciprocal-Gamma identity holds for every $\sigma>0$, we may choose the vertical line separately for this fixed $y$. Set $\sigma=\sqrt{sy}$ and apply the change of variables $w=z\sqrt{y}$, where $z=\sqrt{s}+it$.
Thus $\tau=\sqrt{y}t$ and $\mathrm{d}\tau=\sqrt{y}\mathrm{d}t$. Substituting this change of variables gives
\begin{equation*}
\frac{1}{\Gamma(r+1)} = \frac{1}{\pi} \int_{-\infty}^{\infty} (z\sqrt{y})^{-(2r+1)} e^{y z^2} (\sqrt{y} \mathrm{d}t).
\end{equation*}
Grouping the terms involving $y$ and rearranging provides an integral representation for the deterministic power $y^r$:
$$
y^r = \frac{\Gamma(r+1)}{\pi} \int_{-\infty}^{\infty} z^{-(2r+1)} e^{y z^2} \mathrm{d}t.
$$
This derivation applies away from the poles of $\Gamma(r+1)$; at $r=-1,-2,\ldots$, the identity is understood by taking the limiting value, as stated in the theorem.
To obtain the moment $\mathbb{E}[(X-\xi)^r]$, we substitute $y = X-\xi$ and take the expectation of both sides. 
To justify the interchange of expectation and integration via Fubini's theorem, let
$Y=X-\xi>0$ almost surely and set $p=\operatorname{Re}(r)$. With
$z=\sqrt{s}+it$, the kernel satisfies $|e^{Yz^2}|=e^{Y(s-t^2)}$, and the absolute integral is bounded by a constant depending only on $r$ times
\begin{equation*}
\int_{-\infty}^{\infty}
|z|^{-(2p+1)}
\mathbb{E}\left[e^{Y(s-t^2)}\right]\mathrm{d}t.
\end{equation*}
For $p>0$, absolute convergence follows immediately from
$\mathbb{E}[e^{sY}]<\infty$ and the integrability of
$|z|^{-(2p+1)}$ over $t\in\mathbb{R}$.

Now consider the case $p<0$. The integral over compact subsets of
$\mathbb{R}$ is finite because $s>0$ and $\mathbb{E}[e^{sY}]<\infty$.
For the tails, take $|t|>\sqrt{s+1}$ and set $c=t^2-s>1$. Then
$t^2=c+s$, $|z|^2=s+t^2=c+2s$, and
$\mathrm{d}t=\{2(c+s)^{1/2}\}^{-1}\mathrm{d}c$. Hence the two tails are
bounded by a constant times
\begin{equation*}
\int_1^\infty c^{-p-1}\mathbb{E}\left[e^{-Yc}\right]\mathrm{d}c
\leq
\int_0^\infty c^{-p-1}\mathbb{E}\left[e^{-Yc}\right]\mathrm{d}c.
\end{equation*}
By Tonelli's theorem and the change of variables $u=Yc$,
\begin{equation*}
\int_0^\infty c^{-p-1}\mathbb{E}\left[e^{-Yc}\right]\mathrm{d}c
=
\mathbb{E}\left[\int_0^\infty c^{-p-1}e^{-Yc}\mathrm{d}c\right]
=
\Gamma(-p)\mathbb{E}[Y^p].
\end{equation*}
This quantity is finite by the stated assumption
$\mathbb{E}[(X-\xi)^p]<\infty$. Thus the Fubini interchange is justified
for all $p<0$ for which the corresponding negative moment exists.

It remains to cover the boundary case $p=0$. The integral over compact
sets is again finite. For the tails, the same change of variables gives a
bound by a constant times
\begin{equation*}
\int_1^\infty c^{-1}\mathbb{E}\left[e^{-Yc}\right]\mathrm{d}c
=
\mathbb{E}\left[\int_1^\infty c^{-1}e^{-Yc}\mathrm{d}c\right].
\end{equation*}
The inner integral is the exponential integral $E_1(Y)$, and it is bounded
by a constant multiple of $1+|\log Y|$. Hence the boundary case is justified
under the stated logarithmic condition
$\mathbb{E}\left[1+\left|\log(X-\xi)\right|\right]<\infty$.

Consequently,
\begin{equation*}
\mathbb{E}[(X-\xi)^r]
=
\frac{\Gamma(r+1)}{\pi}
\int_{-\infty}^{\infty}
z^{-(2r+1)}
\mathbb{E}\left[e^{(X-\xi)z^2}\right]\mathrm{d}t
=
\frac{\Gamma(r+1)}{\pi}
\int_{-\infty}^{\infty}
\frac{e^{-\xi z^2}M_X(z^2)}{z^{2r+1}}\mathrm{d}t.
\end{equation*}
This completes the proof.
\hfill{}$\square$\medskip{}

\noindent \textbf{Proof of Corollary \ref{cor:parabolic-negative-integers}.}
Set $q=-r$. For $y>0$, 
the parabolic identity gives
$$
y^{-q}=\frac{1}{\pi} \frac{J(q;y)}{1/\Gamma(1-q)},\quad J(q;y)\equiv\int_{-\infty}^{\infty}z^{2q-1}e^{yz^2}\mathrm{d}t.
$$
For $q\notin{1,2,\ldots}$, the preceding identity implies $J(q;y)=\pi y^{-q}/\Gamma(1-q)$; hence $J(q;y)$ extends continuously to $q=m$ and satisfies $J(m;y)=0$ because $1/\Gamma(1-m)=0$.

At $q=m=1,2,\ldots$, both $J(q;y)$ and $1/\Gamma(1-q)$ vanish. Applying l'Hôpital's rule gives
$$
y^{-m}
=
\frac{1}{\pi} \frac{J^\prime(m;y)}{(-1)^m(m-1)!}=
\frac{2(-1)^m}{\pi(m-1)!}
\int_{-\infty}^{\infty}\log(z)z^{2m-1}e^{yz^2}\mathrm{d}t,
$$
since
$
J^\prime(m;y) =
2\int_{-\infty}^{\infty}\log(z)z^{2m-1}e^{yz^2}\mathrm{d}t$.
Setting $y=X-\xi$ and taking expectations gives the stated formula. The additional logarithmic integrability condition justifies the interchange of expectation and integration.
\hfill{}$\square$

\setcounter{equation}{0}
\global\long\def\theequation{S.\arabic{equation}}%

\setcounter{figure}{0}
\global\long\def\thefigure{S.\arabic{figure}}%

\setcounter{lemma}{0}
\global\long\def\thelemma{S.\arabic{lemma}}%

\global\long\def\thecor{S.\arabic{cor}}%
\newpage{}

\setcounter{page}{1}
\global\long\def\thepage{S.\arabic{page}}%

\part*{Supplementary Material}
\renewcommand{\thesection}{S.\arabic{section}}
\renewcommand{\thetable}{S.\arabic{table}}
\renewcommand{\thefigure}{S.\arabic{figure}}
\setcounter{section}{0}
\setcounter{table}{0}
\setcounter{figure}{0}

\section{Auxiliary Contour Identities and Oscillatory Bounds} \label{sec:AuxContourBounds}

\subsection{The vanishing identity}

\noindent \textbf{Detailed Proof of Lemma A.1} 
Consider the function $f(t) = e^{-itx} w^{-(r+1)} = e^{-itx}(\sigma+it)^{-(r+1)}$. For $x > 0$, the term $e^{-itx}$ decays exponentially in the lower half of the complex $t$-plane. We define a closed contour $\mathcal{C}_R$ consisting of the interval $[-R, R]$ on the real axis and a semi-circle of radius $R$ in the lower half-plane.
\begin{figure}[htbp]
\begin{center}
\begin{tikzpicture}[
    scale=0.75,
    midarrow/.style={
        decoration={
            markings, 
            mark=at position 0.62 with {\arrow{>}}
        },
        postaction={decorate}
    }
]

	\coordinate (A) at (-4,0);  
	\coordinate (B) at (4,0);   
	\coordinate (P) at (0,1.0); 
		
	\draw[->, very thick] (-6.2,0) -- (6.2,0) node[below] {$\color{black}\text{Re}(t)$}; 
	\draw[->, very thick] (0,-5.0) -- (0,2.5) node[above] {$\color{black}\text{Im}(t)$};
	
	\node[below] at (4.2,0) {$R$}; 
	\node[below] at (-4.4,0) {$-R$}; 
	
	\draw[thick, blue, midarrow] (A) -- (B);
    
	\draw[thick, blue, midarrow] (B) arc (0:-180:4);

	\node at (2.5,-2.5) {${\color{blue}C_{R}}$};

	\fill[white] (P) circle (3pt); 
	\node[teal, right] at (P) {\ Singularity}; 
    \node[left, teal] at (P) {$i\sigma$}; 
\draw[teal, line width=2pt] (P) circle (2.5pt);
\end{tikzpicture} 
\label{fig:contour}
\end{center}
\caption{The integration contour $C_R$ in the complex $t$-plane. The path along the real axis $[-R, R]$ is closed by a semi-circle in the lower half-plane. The singularity at $t=i\sigma$ lies outside the contour.}
\end{figure}
Write $t=u+iv$ with $v\le 0$ on $C_R$ and set $w=\sigma+it$. 
The only potential singularity of the integrand arises at $w=0$, equivalently at $t=i\sigma$.
This point lies in the upper half-plane and is therefore outside $\mathcal{C}_R$.
Moreover, for $t=u+iv$ with $v\le 0$ on (and inside) $\mathcal{C}_R$ we have
$
w=\sigma+it=\sigma+i(u+iv)=\sigma+iu-v$,
$\operatorname{Re}(w)=\sigma-v\ge\sigma>0$,
so the contour lies entirely in the half-plane $\operatorname{Re}(w)>0$ and hence avoids the
principal branch cut of $\log(w)$ on $(-\infty,0]$. Consequently, the integrand is analytic on and
inside $\mathcal{C}_R$, and by Cauchy's theorem,
$\oint_{\mathcal{C}_R} f(t)dt=0$. 

As $R \to \infty$, the integral along the circular arc vanishes due to the exponential decay of $e^{-itx}$ (Jordan's Lemma) and the algebraic decay of $|w|^{-(r+1)}$. Therefore, the integral along the real axis must also be zero.
\hfill{}$\square$\medskip{}

\subsection{Oscillatory bounds}

\begin{lemma}[Oscillatory bound]
\label{lem:osc-bound}
Let $s>0$, $z=s+it$, and let $p=\operatorname{Re}(r)\in(-1,0)$. For each fixed $r$ there exists a constant $C_r<\infty$ such that, for every $x\neq0$,
$$
\sup_{T>0}
\left|
\int_{-T}^{T}
\frac{e^{zx}+e^{-zx}}{z^{r+1}}\mathrm{d}t
\right|
\leq
C_r(e^{sx}+e^{-sx})|x|^p.
$$
\end{lemma}

\noindent \textbf{Proof.}
It is enough to consider one of the two exponential terms. For the term involving $e^{zx}$, factor out the real exponential and write
$$
\int_{-T}^{T}\frac{e^{(s+it)x}}{(s+it)^{r+1}}\mathrm{d}t
=
e^{sx}I_T(x|r),
\qquad
I_T(x|r)
=
\int_{-T}^{T}\frac{e^{itx}}{(s+it)^{r+1}}\mathrm{d}t.
$$
Set $u=t|x|$, $S=T|x|$, and $a=s|x|$. Then
$$
I_T(x|r)
=
|x|^r
\int_{-S}^{S}
\frac{e^{iu\operatorname{sgn}(x)}}{(a+iu)^{r+1}}\mathrm{d}u.
$$
It is therefore sufficient to show that, for a constant depending only on $r$,
$$
\sup_{S>0,a>0}
\left|
\int_0^S
\frac{e^{\pm iu}}{(a\pm iu)^{r+1}}\mathrm{d}u
\right|
<\infty.
$$

Let $U>0$ be fixed. On the interval $[0,U]$, we have
$$
\left|
\int_0^U
\frac{e^{\pm iu}}{(a\pm iu)^{r+1}}\mathrm{d}u
\right|
\leq
e^{\pi|\operatorname{Im}(r)|}
\int_0^U
u^{-p-1}\mathrm{d}u
=
\frac{e^{\pi|\operatorname{Im}(r)|}U^{-p}}{-p},
$$
which is finite because $p<0$.

For the interval $[U,S]$, assuming $S>U$, integration by parts gives
$$
\int_U^S
\frac{e^{\pm iu}}{(a\pm iu)^{r+1}}\mathrm{d}u
=
\left[
\pm
\frac{e^{\pm iu}}{i(a\pm iu)^{r+1}}
\right]_U^S
+
(r+1)
\int_U^S
\frac{e^{\pm iu}}{(a\pm iu)^{r+2}}\mathrm{d}u.
$$
The remainder is bounded by
$$
\int_U^S
\left|
\frac{1}{(a\pm iu)^{r+2}}
\right|\mathrm{d}u
\leq
e^{\pi|\operatorname{Im}(r)|}
\int_U^S u^{-p-2}\mathrm{d}u
\leq
\frac{e^{\pi|\operatorname{Im}(r)|}}{p+1}U^{-p-1},
$$
because $p>-1$. The boundary terms are also bounded by a constant depending only on $r$ and $U$. Hence
$$
\sup_{S>0,a>0}
\left|
\int_0^S
\frac{e^{\pm iu}}{(a\pm iu)^{r+1}}\mathrm{d}u
\right|
\leq C_r<\infty.
$$
Consequently,
$$
|I_T(x|r)|
\leq
C_r |x|^p.
$$
Multiplying by the outer factor $e^{sx}$ gives the bound for the term involving $e^{zx}$. The term involving $e^{-zx}$ is obtained by replacing $x$ with $-x$, and gives the factor $e^{-sx}$. Combining the two terms yields
$$
\sup_{T>0}
\left|
\int_{-T}^{T}
\frac{e^{zx}+e^{-zx}}{z^{r+1}}\mathrm{d}t
\right|
\leq
C_r(e^{sx}+e^{-sx})|x|^p.
$$
This completes the proof.
\hfill{}$\square$\medskip{}

\begin{lemma}[Boundary oscillatory bound]
\label{lem:boundary-osc-bound}
Let $s>0$, $z=s+it$, and let $r=i\eta$ with $\eta\in\mathbb{R}$. For each fixed $\eta$ there exists a constant $C_{\eta}<\infty$ such that, for every $x\neq0$,
$$
\sup_{T>0}
\left|
\int_{-T}^{T}
\frac{e^{zx}+e^{-zx}}{z^{1+i\eta}}\mathrm{d}t
\right|
\leq
C_{\eta}(e^{sx}+e^{-sx}).
$$
For $\eta=0$, the same conclusion holds with $C_0<\infty$.
\end{lemma}

\noindent \textbf{Proof.}
It is enough to bound one exponential term. Fix $\varepsilon\in\{-1,1\}$ and consider
$$
K_{\varepsilon,T}(x)
=
\int_{-T}^{T}
\frac{e^{\varepsilon z x}}{z^{1+i\eta}}\mathrm{d}t.
$$
Set $u=t|x|$, $S=T|x|$, and $a=s|x|$. Since
$z=s+it=(a+iu)/|x|$, we obtain
$$
K_{\varepsilon,T}(x)
=
e^{\varepsilon sx}|x|^{i\eta}
\int_{-S}^{S}
\frac{e^{i\varepsilon u\operatorname{sgn}(x)}}{(a+iu)^{1+i\eta}}\mathrm{d}u.
$$
Because $|x|^{i\eta}$ has modulus one, it remains to show that
$$
\sup_{S>0,a>0}
\left|
\int_{-S}^{S}
\frac{e^{i\varepsilon u}}{(a+iu)^{1+i\eta}}\mathrm{d}u
\right|
<\infty,
$$
with a constant depending only on $\eta$.

Fix $U>0$ and split the integral into the central part $|u|\leq U$ and the tails $|u|>U$. For the central part, write
$$
e^{i\varepsilon u}=1+\{e^{i\varepsilon u}-1\}.
$$
First suppose $\eta\neq0$. Since
$$
\frac{\mathrm{d}}{\mathrm{d}u}(a+iu)^{-i\eta}
=
\eta(a+iu)^{-1-i\eta},
$$
we have, for $0<B\leq U$,
$$
\left|
\int_{-B}^{B}(a+iu)^{-1-i\eta}\mathrm{d}u
\right|
\leq
\frac{2e^{|\eta|\pi/2}}{|\eta|}.
$$
For $\eta=0$,
$$
\int_{-B}^{B}\frac{1}{a+iu}\mathrm{d}u
=
\frac{1}{i}\{\log(a+iB)-\log(a-iB)\},
$$
and this is bounded uniformly in $a>0$ and $B>0$. The remaining central term is also uniformly bounded, since $|e^{i\varepsilon u}-1|\leq |u|$ and
$$
\int_{-B}^{B}
\frac{|u|}{|a+iu|}
\mathrm{d}u
\leq
2B
\leq
2U,
$$
up to the harmless phase factor coming from $(a+iu)^{-i\eta}$. Thus the contribution from $|u|\leq U$ is bounded uniformly in $S$ and $a$.

It remains to bound the tails. On $[U,S]$, integration by parts gives
$$
\int_U^S
\frac{e^{i\varepsilon u}}{(a+iu)^{1+i\eta}}\mathrm{d}u
=
\left[
\frac{e^{i\varepsilon u}}{i\varepsilon(a+iu)^{1+i\eta}}
\right]_U^S
+
\frac{1+i\eta}{\varepsilon}
\int_U^S
\frac{e^{i\varepsilon u}}{(a+iu)^{2+i\eta}}\mathrm{d}u.
$$
Since $|a+iu|\geq u$ for $u>0$, the boundary terms are bounded by a constant times $U^{-1}$, and the remainder is bounded by a constant times
$$
\int_U^\infty u^{-2}\mathrm{d}u<\infty.
$$
The same argument applies on $[-S,-U]$. Therefore
$$
\sup_{S>0,a>0}
\left|
\int_{-S}^{S}
\frac{e^{i\varepsilon u}}{(a+iu)^{1+i\eta}}\mathrm{d}u
\right|
\leq C_\eta<\infty.
$$
Consequently,
$$
\sup_{T>0}
\left|
\int_{-T}^{T}
\frac{e^{zx}}{z^{1+i\eta}}\mathrm{d}t
\right|
\leq
C_\eta e^{sx},
\qquad
\sup_{T>0}
\left|
\int_{-T}^{T}
\frac{e^{-zx}}{z^{1+i\eta}}\mathrm{d}t
\right|
\leq
C_\eta e^{-sx}.
$$
Adding the two bounds gives
$$
\sup_{T>0}
\left|
\int_{-T}^{T}
\frac{e^{zx}+e^{-zx}}{z^{1+i\eta}}\mathrm{d}t
\right|
\leq
C_\eta(e^{sx}+e^{-sx}).
$$
The proof includes the case $\eta=0$ by the separate central bound above.
\hfill{}$\square$\medskip{}

The proofs of Lemmas \ref{lem:osc-bound} and \ref{lem:boundary-osc-bound} also give the corresponding one-sided estimates
$$
\sup_{T>0}
\left|
\int_{-T}^{T}
\frac{e^{zx}}{z^{r+1}}\mathrm{d}t
\right|
\leq
C_r e^{sx}|x|^{\operatorname{Re}(r)},
\qquad -1<\operatorname{Re}(r)<0,
$$
and
$$
\sup_{T>0}
\left|
\int_{-T}^{T}
\frac{e^{zx}}{z^{1+i\eta}}\mathrm{d}t
\right|
\leq
C_\eta e^{sx},
\qquad \eta\in\mathbb{R}.
$$

\begin{lemma}[Logarithmic oscillatory bound]
\label{lem:log-osc-bound}
Let $s>0$, $z=s+it$, and let $p=\operatorname{Re}(r)\in(-1,0]$. For each fixed $r$ there exists a constant $C_{r,s}<\infty$ such that, for every $x\neq0$,
$$
\sup_{T>0}
\left|
\int_{-T}^{T}
\frac{e^{zx}+e^{-zx}}{z^{r+1}}\log(z)\mathrm{d}t
\right|
\leq
C_{r,s}(e^{sx}+e^{-sx})|x|^p(1+|\log|x||),
$$
where $\log(z)$ denotes the principal logarithm.
\end{lemma}

\noindent \textbf{Proof.}
It is enough to consider one of the two exponential terms. For the term involving $e^{zx}$, write
$$
\int_{-T}^{T}
\frac{e^{zx}}{z^{r+1}}\log(z)\mathrm{d}t
=
e^{sx}
\int_{-T}^{T}
\frac{e^{itx}}{(s+it)^{r+1}}\log(s+it)\mathrm{d}t.
$$
Set $u=t|x|$, $S=T|x|$, and $a=s|x|$. Since
$s+it=(a+iu)/|x|$, we have
$$
\log(s+it)=\log(a+iu)-\log|x|,
$$
and therefore
$$
\int_{-T}^{T}
\frac{e^{itx}}{(s+it)^{r+1}}\log(s+it)\mathrm{d}t
=
|x|^r
\int_{-S}^{S}
\frac{e^{iu\operatorname{sgn}(x)}}{(a+iu)^{r+1}}
\{\log(a+iu)-\log|x|\}\mathrm{d}u.
$$

First suppose $p\in(-1,0)$. The term involving $-\log|x|$ is controlled by Lemma \ref{lem:osc-bound} and contributes at most a constant times
$$
|x|^p|\log|x||.
$$
It remains to bound the term with $\log(a+iu)$. The same splitting and integration-by-parts argument used in Lemma \ref{lem:osc-bound} applies, with the logarithmic factor carried through the estimates. On bounded intervals, the integrand is bounded by an integrable envelope of the form
$$
|u|^{-p-1}(1+|\log|u||),
$$
away from $u=0$, while the region near $u=0$ contributes at most a constant times $1+|\log a|$. Since $a=s|x|$, this term is bounded by a constant times $1+|\log|x||$, with the constant allowed to depend on $s$. On the tails, integration by parts gives boundary terms of order
$$
(1+\log u)u^{-p-1}
$$
and a remainder bounded by
$$
\int_U^\infty (1+\log u)u^{-p-2}\mathrm{d}u<\infty,
\qquad p>-1.
$$
Thus, for $p\in(-1,0)$, the logarithmic term is bounded by
$$
C_{r,s}|x|^p(1+|\log|x||).
$$

Now consider the boundary case $p=0$, so $r=i\eta$ with $\eta\in\mathbb{R}$. In the decomposition above, the term involving $-\log|x|$ is controlled by Lemma \ref{lem:boundary-osc-bound} and contributes at most a constant times $|\log|x||$. It remains to bound the term with $\log(a+iu)$.

Fix $\varepsilon\in\{-1,1\}$ and write
$$
J_{\varepsilon,S,a}
=
\int_{-S}^{S}
\frac{e^{i\varepsilon u}}{(a+iu)^{1+i\eta}}
\log(a+iu)\mathrm{d}u.
$$
We show that
$$
\sup_{S>0}|J_{\varepsilon,S,a}|
\leq
C_{\eta,s}(1+|\log a|).
$$
Since $a=s|x|$, this gives the desired factor $1+|\log|x||$.

Fix $U=1$ and first consider the central part $|u|\leq U$. Write
$e^{i\varepsilon u}=1+\{e^{i\varepsilon u}-1\}$. For the term with $1$, if
$\eta\neq0$, an antiderivative is
$$
\frac{1}{\eta}(a+iu)^{-i\eta}
\left\{\log(a+iu)+\frac{1}{i\eta}\right\}.
$$
Thus the integral over any interval $[-B,B]$, $0<B\leq U$, is bounded by
$C_\eta(1+|\log a|)$. If $\eta=0$, the corresponding antiderivative is
$\frac{1}{2i}\{\log(a+iu)\}^2$. 
Evaluating between $-B$ and $B$ again gives a bound of order
$C(1+|\log a|)$, because the imaginary part of $\log(a+iB)$ is uniformly
bounded. For the remaining central term, $|e^{i\varepsilon u}-1|\leq |u|$, and
$$
\int_{-U}^{U}
\frac{|u|}{|a+iu|}
|\log(a+iu)|\mathrm{d}u
\leq
C(1+|\log a|).
$$
Hence the central part is bounded by $C_\eta(1+|\log a|)$.

It remains to bound the tails. On $[U,S]$, integration by parts gives
$$
\int_U^S
\frac{e^{i\varepsilon u}\log(a+iu)}{(a+iu)^{1+i\eta}}\mathrm{d}u
=
\left[
\frac{e^{i\varepsilon u}\log(a+iu)}
{i\varepsilon(a+iu)^{1+i\eta}}
\right]_U^S
-
\frac{1}{i\varepsilon}
\int_U^S
e^{i\varepsilon u}
\frac{\mathrm{d}}{\mathrm{d}u}
\left\{\frac{\log(a+iu)}{(a+iu)^{1+i\eta}}\right\}
\mathrm{d}u.
$$
The boundary terms are uniformly bounded in $S$ and $a$, since
$|a+iu|\geq U$ and $(1+|\log v|)/v$ is bounded for $v\geq U$. Moreover,
$$
\left|
\frac{\mathrm{d}}{\mathrm{d}u}
\left\{\frac{\log(a+iu)}{(a+iu)^{1+i\eta}}\right\}
\right|
\leq
C_\eta
\frac{1+|\log(a+iu)|}{|a+iu|^2}.
$$
The integral of the right-hand side over $[U,\infty)$ is bounded uniformly in
$a>0$. To see this, if $a\leq U$ then $|a+iu|\geq u$ and the envelope is bounded
by $C(1+\log u)u^{-2}$. If $a>U$, split the integral into $[U,a]$ and
$[a,\infty)$; the first part is bounded by $C(1+\log a)/a$, and the second by
$C\int_a^\infty(1+\log u)u^{-2}\mathrm{d}u$. Both are uniformly bounded. The
same argument applies on the negative tail $[-S,-U]$.

Combining the central and tail bounds gives
$$
\sup_{S>0}|J_{\varepsilon,S,a}|
\leq
C_{\eta,s}(1+|\log a|)
\leq
C_{\eta,s}(1+|\log|x||).
$$
Multiplying by the outer factor $e^{sx}$ gives the desired bound for the term
with $e^{zx}$. The term with $e^{-zx}$ follows by replacing $x$ with $-x$.
\hfill{}$\square$\medskip{}

\section{Standardized NIG Parameter Mappings}
\label{app:nig_mappings}

For the numerical illustrations in Section 3, we utilize standardized normal-inverse Gaussian (NIG) distributions (zero mean and unit variance). Following \citet{BNBJS1985nig}, these can be characterized entirely by two shape parameters, $(\xi,\chi)$, satisfying $0\leq|\chi|<\xi<1$. The exact algebraic mapping to the standard parameters $(\alpha, \beta, \mu, \delta)$ is given by:
$$
 \alpha=\xi\zeta, \quad \beta=\chi\zeta,\quad \mu=-\frac{\chi(\xi^2-\chi^2)}{\xi^2}\zeta, \quad \delta=\frac{(\xi^2-\chi^2)^{3/2}}{\xi^2}\zeta, \quad \text{where }\zeta=\frac{\sqrt{1-\xi^{2}}}{\xi^{2}-\chi^{2}}.
$$
For the absolute-moment formula in Theorem \ref{thm:factional-abs-moments}, the contour shift must be such that both $M_X(s+it)$ and $M_X(-s-it)$ are finite. Hence the relevant condition is $0<s<\alpha-|\beta|$. For the two standardized distributions considered in the main text, these upper limits are $3\sqrt{3}/5\approx1.04$ and $2\sqrt{7}\approx5.29$, respectively, so the choice $s=1$ is admissible in both cases.

\section{Integrand Behavior for the NIG Distribution}
\label{app:nig_integrands}

\begin{figure}[htb]
\centering{}\includegraphics[width=1\textwidth]{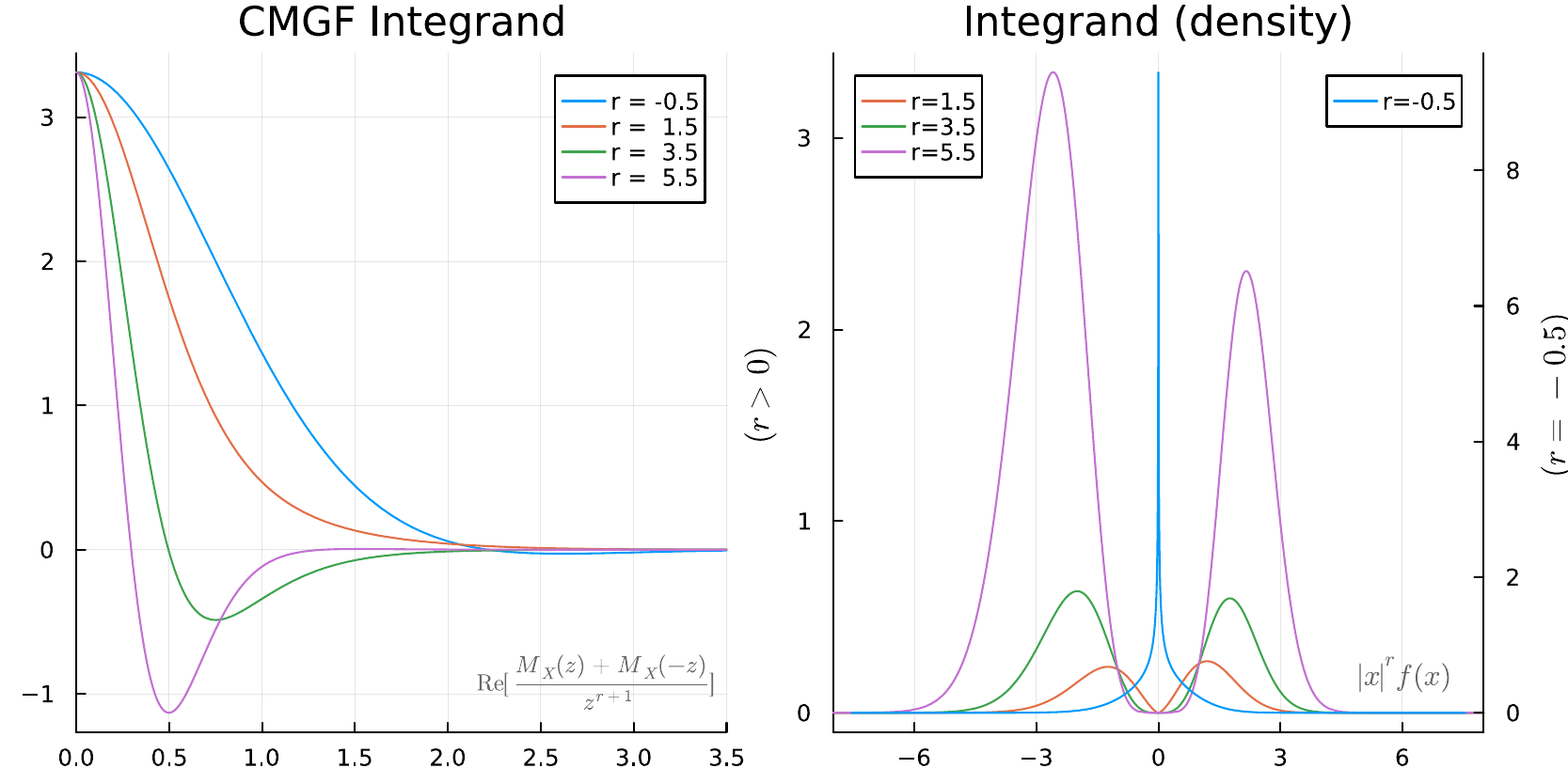}
\caption{{\small Integrand profiles for selected moments of the NIG distribution. The CMGF integrands (left) remain stable across fractional powers $r$, whereas the density-based integrands (right) vary substantially in scale and effective domain.\label{fig:NIG-Integrands}}}
\end{figure}

The shape of the integrands provides visual intuition for the computational stability of the CMGF method relative to classical density integration. Figure \ref{fig:NIG-Integrands} plots the corresponding integrands for the NIG distribution with $(\xi,\chi)=(1/2,-1/3)$.

The CMGF integrand (evaluating the complex contour) is very stable: it is non-negligible over a nearly identical domain for all values of $r$, and the range of the integrand is completely consistent across $r$. In fact, all CMGF integrands share the exact same maximum value at zero. By contrast, the classical integrand, $|x|^{r}f(x)$, exhibits severe structural variation across different moments, explaining why classical quadrature methods struggle and slow down significantly for higher fractional powers.

\section{Existing Integral Expressions for Fractional Moments}
\label{app:existing_methods}

To contextualize the theoretical contribution of the CMGF framework, we summarize the existing integral expressions for fractional moments, which predominantly rely on characteristic functions (CF) or fractional derivatives rather than the complex MGF.

For positive fractional moments, \citet[Theorem 11.4.4]{Kawata1972} established:
$$
\mathbb{E}|X|^{r}=C_{K}\int_{0}^{\infty}u^{-(1+r)}\left[-\operatorname{Re}[\varphi_{X}(u)]+\sum_{k=0}^{K}\frac{u^{2k}}{(2k)!}\varphi_{X}^{(2k)}(0)\right]\mathrm{d}u,
$$
where $K=\lfloor r/2 \rfloor$ and $C_{K}$ is a positive constant. Similarly, \citet{Laue1980} derived:
$$
\mathbb{E}|X|^{r} =\frac{\lambda}{\Gamma(1-\lambda)}[\cos(\tfrac{\pi r}{2})]^{-1}\int_{0}^{\infty}\frac{\varphi_{X}^{(k)}(0)-\varphi_{X}^{(k)}(u)}{u^{1+\lambda}}\mathrm{d}u,
$$
where $k=\lfloor r\rfloor$ and $\lambda=r-k$. 

For strictly non-negative random variables ($X\geq0$), \citet{SchurgerK:2002} provided an expression closely related to Laue's:
$$
\mathbb{E}[X^{r}]=(-1)^{k}\frac{\lambda}{\Gamma(1-\lambda)}\int_{0}^{\infty}\frac{M_{X}^{(k)}(0)-M_{X}^{(k)}(-u)}{u^{1+\lambda}}\mathrm{d}u.
$$
An alternative expression for non-negative variables utilizing higher-order derivatives was established by \citet{CressieBorkent1986}:
$$
\mathbb{E}[X^{r}] =\frac{1}{\Gamma(\tilde{\lambda})}\int_{0}^{\infty}u^{\tilde{\lambda}-1}M_{X}^{(\tilde{k})}(-u)\mathrm{d}u,
$$
where $\tilde{k}=\lceil r\rceil>0$ and $\tilde{\lambda}=\tilde{k}-r\in[0,1)$. Finally, for strictly positive variables ($X>0$) and negative fractional moments ($r<0$), \citet[Theorem 1.1]{SchurgerK:2002} established:
$$
\mathbb{E}[X^{r}]=\frac{1}{\Gamma(-r)}\int_{0}^{\infty}\frac{1}{t^{r+1}}M_{X}(-t)\mathrm{d}t.
$$
The CMGF framework introduced in this paper unifies these previously separate strands in the sense that the same Fourier-Laplace contour mechanism yields derivative-free formulas for absolute moments, positive-part moments, signed integer moments, logarithmic moments, complex powers, and negative moments under explicit regularity conditions.

\section{Numerical Implementation and Supplementary Checks}
\label{app:computational_details}

This section provides the computational details for the numerical illustrations in Section 3, the timing comparison for the NIG example, and a supplementary numerical check of the parabolic representation in Theorem \ref{thm:parabolic-positive}.

\subsection{Hardware and Software Environment}
All numerical evaluations, integrations, and Monte Carlo simulations were executed using the Julia programming language, version 1.11.0 \citep{Julia2017}. Computation times were rigorously evaluated using the \texttt{BenchmarkTools.jl} package \citep{BenchmarkTools.jl-2016}, which executes multiple samples to provide statistically robust minimum, median, and mean execution times. The algorithms were run on an Apple MacBook Pro equipped with an M1 Max processor and 32 GB of unified memory. 

\subsection{Numerical Integration Tolerances}
To ensure a fair comparison in Table \ref{tab:NIGmoments}, both the CMGF integral (evaluated via Theorem \ref{thm:factional-abs-moments}) and the classical density-based expectation, $\int_{-\infty}^{\infty}|x|^{r}f(x)\mathrm{d}x$, were computed using the same adaptive numerical quadrature routines in Julia. Both methods were subject to identical relative and absolute error tolerance thresholds (e.g., \texttt{rtol = 1e-10}). The computational stability of the CMGF integral is highly insensitive to the specific choice of $s$, provided it falls strictly within the convergence strip.

\subsection{Monte Carlo Simulation Setup}
The simulation-based moment estimates ($\hat{\mu}_{N}^{r}$) reported in Section 3 were computed as the empirical average of $N=1,000,000$ independent and identically distributed random draws from the respective NIG distributions:
$$
\hat{\mu}_{N}^{r} = \frac{1}{N}\sum_{j=1}^{N}|X_{j}|^{r}.
$$
To accurately quantify the simulation error and determine the number of reliable decimal places plotted in Figure \ref{fig:MomentsNIG}, we estimated the variance of the fractional powers, $\sigma^{2}(r)=\operatorname{var}(|X_{j}|^{r})$, using an extended auxiliary simulation of $100,000,000$ draws. The number of accurate decimal places corresponding to a one-standard-deviation error was then calculated as $-\log_{10}(\sigma(r)/\sqrt{N})$.

\subsection{Timing comparison for the NIG example}

Table \ref{tab:NIGmoments} reports the computation times for the NIG example in Section \ref{sec:applications}. We compare four approaches: the CMGF method of Theorem \ref{thm:factional-abs-moments}, differentiation of the MGF for the integer moments $r=2$ and $r=4$, direct numerical integration of the density, and Monte Carlo simulation based on $N=1,000,000$ draws. The CMGF method is substantially faster than density-based quadrature in this example because it evaluates the closed-form MGF directly, whereas the density-based method requires repeated evaluations of modified Bessel functions.

\begin{table}[th]
\caption{CMGF Computation Time of $\mathbb{E}|X|^{r}$ for $X\sim\operatorname{NIG}$}
\begin{centering}
\vspace{0.2cm}
\begin{small}
\begin{tabularx}{\textwidth}{YYYYY}
\toprule
\midrule
  $r$ &  CMGF & $M_X^{(k)}$ & Integrate with density & Simulations $(N=10^6)$  \\
\midrule
\\[-0.2cm]    
-0.5 & 28.3 &       & 712.9   &   397,666  \\
 0.5 & 29.2 &       & 244.0   &   395,837  \\
 1.0 & 20.3 &       & 199.1   &   299,232  \\
 1.5 & 29.3 &       & 175.0   &   399,173  \\
 2.0 & 20.4 &  55.7 & 108.3   &   298,635  \\
 2.5 & 24.5 &       & 149.3   &   399,182  \\
 3.0 & 17.0 &       & 122.5   &   291,924  \\
 3.5 & 24.5 &       & 114.7   &   402,552  \\
 4.0 & 17.6 & 104.9 & 100.3   &   318,081  \\
\\[-0.2cm]
\midrule
\bottomrule
\end{tabularx}
\end{small}
\par\end{centering}
{\small Note: Computation time in microseconds (\textmu s) for evaluating $\mathbb{E}|X|^{r}$ using four methods: the CMGF method of Theorem \ref{thm:factional-abs-moments}, the $k$-th derivative of the MGF (for $r=2$ and $r=4$), numerical integration of $\int_{-\infty}^{\infty}|x|^{r}f(x)\mathrm{d}x$, and standard Monte Carlo simulation ($N=1,000,000$). Both CMGF and the density-based method use numerical integration with identical tolerance thresholds. Hardware and software environment details are provided in the Supplementary Material.\label{tab:NIGmoments}}{\small\par}
\end{table}

\subsection{Numerical check of the parabolic representation}

As a numerical check of Theorem \ref{thm:parabolic-positive}, we consider $X\sim\chi^2_8$. Its MGF is $M_X(z)=(1-2z)^{-4}$ for $\operatorname{Re}(z)<1/2$, and its fractional moments are available in closed form:
$$
\mathbb{E}[X^r]=2^r\frac{\Gamma(4+r)}{\Gamma(4)},\qquad r>-4.
$$
Using the parabolic contour $z=\sqrt{s}+it$ with $s=0.1$, the integral in Theorem \ref{thm:parabolic-positive} matches the closed-form expression to machine precision over a wide range of positive and negative powers. This includes negative fractional moments close to the existence boundary $r=-4$.

We also evaluate the pole cases $r=-1,-2,-3$. Although the formula in Theorem \ref{thm:parabolic-positive} contains the factor $\Gamma(r+1)$, the right-hand side is understood in the limiting sense at $r=-1,-2,\ldots$. The rows labeled ``limit'' in Table \ref{tab:chisq_parabolic_check} evaluate the theorem slightly away from the poles, at $r=-m+10^{-6}$. The rows labeled ``log'' use the pole-free logarithmic contour formula from Corollary \ref{cor:parabolic-negative-integers} at the exact negative integers. For $X\sim\chi^2_8$, the corresponding moments are $\mathbb{E}[X^{-1}]=1/6$, $\mathbb{E}[X^{-2}]=1/24$, and $\mathbb{E}[X^{-3}]=1/48$.

\begin{table}[ht]
\begin{centering}
\caption{Numerical verification of the parabolic representation for $X\sim\chi^2_8$ with $s=0.1$.}
\label{tab:chisq_parabolic_check}
\begin{footnotesize}
\begin{tabularx}{\textwidth}{
@{\extracolsep{\fill}}
l
l
S[table-format=1.15e-2]
S[table-format=1.15e-2]
S[table-format=1.3e-2]
@{}
}
\toprule
{$\quad r$} & {Method} & {Parabolic integral} & {Exact moment} & {Relative error} \\
\midrule
$-3.5$             & limit & 2.611071119407293e-2 & 2.611071119407292e-2 & 1.329e-16 \\
$-3$               & log   & 2.083333333333322e-2 & 2.083333333333333e-2 & 5.329e-15 \\
$-3+10^{-6}$       & limit & 2.083333574873437e-2 & 2.083333574859052e-2 & 6.905e-12 \\
$-2.5$             & limit & 2.611071119407292e-2 & 2.611071119407292e-2 & 0.000e0 \\
$-2$               & log   & 4.166666666666709e-2 & 4.166666666666666e-2 & 1.033e-14 \\
$-2+10^{-6}$       & limit & 4.166671316371828e-2 & 4.166671316385254e-2 & 3.222e-12 \\
$-1.5$             & limit & 7.833213358221879e-2 & 7.833213358221877e-2 & 1.772e-16 \\
$-1$               & log   & 1.666666666666667e-1 & 1.666666666666667e-1 & 1.665e-16 \\
$-1+10^{-6}$       & limit & 1.666669359805728e-1 & 1.666669359888365e-1 & 4.958e-11 \\
$-0.5$             & limit & 3.916606679110938e-1 & 3.916606679110939e-1 & 2.835e-16 \\
$\hphantom{-}0.5$  & limit & 2.741624675377657e0  & 2.741624675377657e0  & 1.620e-16 \\
$\hphantom{-}1$    & limit & 8.000000000000002e0  & 8.000000000000000e0  & 2.220e-16 \\
$\hphantom{-}2$    & limit & 8.000000000000003e1  & 8.000000000000000e1  & 3.553e-16 \\
$\hphantom{-}3$    & limit & 9.599999999999997e2  & 9.600000000000000e2  & 3.553e-16 \\
\bottomrule
\end{tabularx}
\end{footnotesize}
\par\end{centering}\medskip\par
{\small Note: The numerical results use $s=0.1$, safely inside the admissible range $0<s<1/2$. The rows labeled ``limit'' use Theorem \ref{thm:parabolic-positive}; near the poles of $\Gamma(r+1)$ they evaluate the integral at $r=-m+10^{-6}$. The rows labeled ``log'' use the pole-free logarithmic contour formula of Corollary \ref{cor:parabolic-negative-integers} at the exact negative integers. Values of $s$ closer to the boundary remain valid, but can be less stable near the poles of $\Gamma(r+1)$, where the limiting formula involves cancellation.}
\end{table}

\end{document}